\documentclass[11pt,a4paper]{article}
\pdfoutput=1
\usepackage{jheppub}

\usepackage{amsmath}
\usepackage{amsfonts}
\usepackage{cancel}
\usepackage{booktabs}
\usepackage{multirow}
\usepackage{xcolor}

\newcommand{\order}[1]{\mathcal{O}\!\left(#1\right)}

\newcommand{\derOrd}[2]{\frac{d#1}{d#2}}
\newcommand{\der}[2]{\frac{\partial#1}{\partial#2}}

\newcommand{\non}{\nonumber\\}
\title{Diffusion models for the $O(3)$ nonlinear sigma model}

\author{Daniel Schubring}

\affiliation{Department of Physics, City College of New York, CUNY,\\
	New York, NY 10031, USA}

\emailAdd{daschubring@gmail.com}

\date{}

\abstract{A score-based diffusion model is trained on the two-dimensional classical Heisenberg model, the lattice representation of the $O(3)$ nonlinear sigma model. Neural networks are trained on three different coarse-grainings of the same training set to implement a multiscale sampling scheme that modestly improves both the accuracy and computational cost of direct sampling on the finest lattice. Stein (or Schwinger--Dyson) diagnostics for the two-point correlation function are introduced to assess the quality of the learned score without relying on reverse diffusion sampling. These diagnostics are also incorporated into an additional training loss. We discuss several technical considerations relevant to implementing diffusion models on $S^2$. These include a consideration of lifting the network input to $S^3$, as in the $CP^1$ representation, and developing a boundary-layer approximation to the short-time spherical heat kernel that extends its usefulness to larger diffusion times.}

\begin{document}

	\maketitle

\section{Introduction}

Lattice field theory provides a nonperturbative framework for studying quantum field theories, with physical observables conventionally evaluated using Markov chain Monte Carlo (MCMC) methods. Despite their successes, conventional MCMC algorithms can suffer from long autocorrelation times as the continuum limit is approached, a manifestation of critical slowing down \cite{wolff1990}. In theories with nontrivial topology, transitions between topological sectors may become strongly suppressed, leading to the additional problem of topological freezing \cite{allesEtAl1996,schaeferSommerVirotta2011,luscherSchaefer2011,bonannoBonatiDElia2021,delDebbio2004,flynnEtAl2015,hasenbusch2017,bonannoNadaVadacchino2024}.

To address these issues, generative machine-learning techniques have been investigated as alternatives or complements to conventional MCMC methods  \cite{albergo2019,kanwar2020,boyda2021,boydaEtAl2022,cranmerEtAl2023,lawrence2025}. In particular, score-based diffusion models (DMs) \cite{hyvarinen2005,vincent2011,sohlDickstein2015,songErmon2019,
	ho2020,songErmon2020,song2021,
	songMengErmon2021,dhariwalNichol2021,
	karras2022} have recently been applied to lattice field theories including scalar $\phi^4$ theory \cite{wangAartsZhou2024,ranner2024,vega2026}, two-dimensional $U(1)$ gauge theory \cite{ranner2024,vega2026,zhu2026}, the $XY$ model \cite{jha2026}, frustrated spin glasses \cite{xiChen2026}, and non-Abelian gauge theories \cite{aarts2026,komijani2026}.
	
While diffusion models offer a promising approach to mitigating critical slowing down and topological freezing, their computational cost remains a challenge. Generating configurations requires integrating ordinary or stochastic differential equations involving many evaluations of a learned score function, which can be considerably more expensive than conventional MCMC updates. Here we investigate the tradeoffs between computational cost and accuracy in diffusion models for lattice field theories, using the two-dimensional $O(3)$ nonlinear sigma model \cite{polyakov1975,brezinZinnJustin1976,bergLuscher1981,berg1981dislocations,wolff1989,apostolakis1991}, also known as the classical Heisenberg model \cite{joyce1967,shenkerTobochnik1980}, as a test case.

The $O(3)$ model shares several qualitative features with four-dimensional Yang--Mills theory, including asymptotic freedom, emergence of a mass gap, and topological instantons, making it a useful testing ground for methods developed to study nonperturbative gauge theories (see, e.g., \cite{novikovShifmanVainshteinZakharov1984}). The $O(3)$ model involves an $S^2$-valued field, but it may equivalently be expressed in terms of an $S^3$-valued field with gauge redundancy, in which case it is known as the $CP^1$ model. The latter interpretation admits a large-$N$ generalization, the $CP^{N-1}$ model \cite{dAddaDiVecchiaLuscher1978,eichenherr1978,witten1979instantons}, which exhibits severe topological freezing in conventional Monte Carlo simulations \cite{delDebbio2004,flynnEtAl2015,hasenbusch2017,bonannoNadaVadacchino2024}. In contrast, the $CP^1$ or $O(3)$ model case we study here does have efficient MCMC methods \cite{wolff1989}. This makes it a useful benchmark for assessing diffusion-based sampling methods with potential applications to the higher $CP^{N-1}$ family and the qualitatively related 4D lattice gauge theories.

Using the $O(3)$ model, we investigate two complementary methodological questions motivated by the computational cost of score evaluations. First, we explore superresolution methods and multiscale sampling  \cite{brandtRon2001,endres2015,efthymiouBeachMelko2019,meng2022,shuLiFarimani2023,cotlerRezchikov2023,shiina2024,bauer2024,singha2026,chen2026skild}, see Fig.~\ref{Fig seurat} and Fig.~\ref{Fig superres}. The diffusion model is used to generate coarse-grained configurations on a smaller, cheaper lattice, and then the results are super-resolved to the original lattice. Second, limiting the number of score evaluations may require large integration step sizes, introducing discretization errors that can be comparable to those arising from the neural network imperfectly learning the score. To evaluate the learned score independently of the discretization error, we introduce diagnostics based on Stein identities \cite{liuLeeJordan2016,barpEtAl2019,xuMatsuda2020,grathwohlEtAl2020}, which are also known as Schwinger-Dyson identities in a field-theory context \cite{gottlob1996,shuklaEtAl2026}.

Additional technical challenges in applying diffusion models to the $O(3)$ model arise because its fields take values on the nonlinear manifold $S^2$. Diffusion models can be naturally extended to Riemannian manifolds by replacing the Gaussian transition kernel with the heat kernel associated with the manifold \cite{deBortoli2022,huangEtAl2022,lou2023,jagvaral2024,kanwarVega2026}. However, heat kernels are generally not available in closed form, and practical approximations are needed for numerical calculations. The case of $S^2$ presents particular difficulties that do not arise for Lie group manifolds. As previously noted in the machine-learning literature \cite{zhaoSong2018}, the short-time asymptotic expansion of the spherical heat kernel becomes singular near the antipodal point. We develop a separate boundary-layer expansion to obtain an accurate approximation in this region. A further complication is that $S^2$ has a non-trivial tangent bundle, and we discuss the implications for the choice of neural network architecture.

An outline of the paper is as follows. In Sec.~\ref{Sec O3}, we review the lattice formulation of the $O(3)$ model and also describe the basic setup of our numerical results with diffusion models. In Sec.~\ref{Sec DM}, we review diffusion models on Riemannian manifolds, focusing on practical aspects of evaluating the heat kernel and its score on $S^2$. In Sec.~\ref{Sec Stein}, we introduce Stein diagnostics that probe the accuracy of the learned score in reproducing two-point correlations, and show how these can be incorporated into the training loss to improve short-distance correlations. In Sec.~\ref{Sec superres}, we describe our superresolution methods and present numerical results using DMs trained at three different levels of coarse graining. Finally, in Sec.~\ref{Sec discussion}, we summarize our findings and discuss future directions.

\section{The $O(3)$ model} 
\label{Sec O3}
\subsection{Action}
The lattice action for the $O(3)$ classical Heisenberg model involves a three-component unit vector field $\mathbf{n}_i$ at the lattice site $i$,
\begin{align}
	S[\mathbf{n};\beta] = -\beta \sum_{\langle ij\rangle} \mathbf{n}_i\cdot  \mathbf{n}_j,\label{S lattice}
\end{align}
and the probability density of a given field configuration is $p(\mathbf{n})\propto e^{-S[\mathbf{n};\beta]}$. Particularly relevant in the context of machine learning is the  \emph{score}, $s = \nabla \log p(\mathbf{n})$, where the gradient is taken in field space, which is a product of factors of $S^2$ for each lattice site. The component of score at lattice site $i$ is
\begin{align}
	\mathbf{s}_i=\nabla_{\mathbf{n}_i} \log p(\mathbf{n}) = \beta\left(\mathbf{m}_i - \left(\mathbf{m}_i \cdot \mathbf{n}_i\right)\mathbf{n}_i\right),\label{eq score exact}
\end{align}
where $\mathbf{m}$ is the nearest neighbor field
\begin{align}
	\mathbf{m}_i = \sum_{j:\langle ij\rangle}\mathbf{n}_j.
\end{align}

Since we will consider coarse-grained field configurations, it is useful to briefly discuss renormalization in the $O(3)$ model. In the continuum limit, up to a constant term, the action becomes
\begin{align}
	S = \frac{\beta}{2}\int d^2x\, \left(\partial_\mu\mathbf{n}(x)\right)^2.
	\label{S continuum}
\end{align}
The continuum theory is renormalizable, with a single running coupling $\beta$ whose renormalization group flow can be calculated perturbatively at weak coupling (large $\beta$).

However, this does not imply that coarse graining the lattice model produces an effective action of the same form with a renormalized coupling $\beta'$. We work at $\beta=1.5$, corresponding to a correlation length of approximately 11 lattice spacings, and consider coarse-grained configurations with even shorter correlation lengths. At these scales, the effective lattice action contains additional operators beyond the nearest-neighbor interaction in \eqref{S lattice} (see, e.g., \cite{gottlob1996,kushnirRosenstein1995}). Consequently, the score of the coarse-grained distribution cannot be obtained from \eqref{eq score exact} by simply replacing $\beta$ with $\beta'$, but nevertheless it can be learned by a diffusion model (see Fig.~\ref{Fig correlations16}, discussed further below).

The continuum model \eqref{S continuum} has an equivalent formulation as the $CP^1$ model \cite{dAddaDiVecchiaLuscher1978,eichenherr1978},
\begin{align}
	S = 2\beta \int d^2 x \left(\partial_\mu z^\dagger\cdot \partial_\mu z + \left(z^\dagger\cdot \partial_\mu z\right)^2\right),\label{S CP1}
\end{align}
which is written in terms of a two-component complex unit vector $z$ related to $\mathbf{n}$ through 
\begin{align}
\mathbf{n}= z^\dagger \boldsymbol{\sigma}z,\qquad  z^\dagger z = 1,\label{Def z}
\end{align}
where $\boldsymbol{\sigma}$ indicates the three Pauli matrices. Note that this relation only defines $z$ up to a local phase, and indeed the action \eqref{S CP1} is gauge invariant under the transformation $z(x) \rightarrow e^{i\alpha(x)}z(x)$. This is often made more explicit by replacing the quartic term in the action with an auxiliary $U(1)$ gauge field, although we shall not need this here.

The conventional lattice discretization of the $CP^1$ action \cite{campostrini1992simulation} is not strictly identical to \eqref{S lattice}. However, using \eqref{Def z} a configuration of $S^2$-valued fields $\mathbf{n}_i$ may always be replaced with $S^3$-valued fields $z_i$, which are defined up to a local $U(1)$ phase. We will return to this observation in Sec.~\ref{Sec DM score parameterization}.
	
\subsection{Observables}
\subsubsection{Two-point correlations}
One of the main observables we measure is the two-point correlation function $C(j)=\left\langle\mathbf n_i\cdot\mathbf n_{i+j}\right\rangle.$ By translation invariance this depends only on the spatial vector displacement $j$.

Some other common observables are determined by $C(j)$. The average energy density is
\begin{align}
	E = \left(\beta V\right)^{-1}\langle S\rangle = -2C(1),
\end{align}
where $V$ indicates the total number of lattice sites, and $C$ is written as a function of a scalar integer when the displacement is aligned along the rectangular axes.

The correlation length $m^{-1}$ may be extracted from the Fourier transform $G(p)=\sum_j e^{i p\cdot j}C(j)$. For small $p$, $G(p)$ takes the schematic form
\begin{align}
	G(p)\approx \frac{Z}{\hat{p}^2+m^2},
\end{align}
where $\hat{p}^2 = 4\sum_\mu \sin^2(p_\mu/2)$. Taking $p$ to be the minimum nonzero lattice momentum, we use this expression to define the squared correlation length $m^{-2}$ on a finite lattice
\begin{align}
	m^{-2}= \frac{1}{\hat{p}^2}\left(\frac{G(0)}{G(p)}-1\right).
\end{align}
Note that the quantity $G(0)$ is the magnetic susceptibility $\chi_m$
\begin{align}
	\chi_m = \frac{1}{V}\left\langle \left(\sum_i \mathbf n_i\right)^2\right\rangle = \sum_j C(j) = G(0).
\end{align}

\subsubsection{Topological charge}

In the continuum, the field $\mathbf{n}$ is a map from spatial positions $x\in\mathcal{M}$ to an $S^2$ target space. The topological charge $Q$ counts the number of times this map wraps the two-dimensional manifold $\mathcal{M}$ around $S^2$, with a sign determined by orientation.

On the lattice, an integer topological charge may still be defined, but its definition depends on how the field $\mathbf{n}$ is interpolated between lattice sites. Berg and Lüscher give a geometric construction of the local topological charge $q_i$ at plaquette $i$ \cite{bergLuscher1981}. Labeling the four vertices of a plaquette consecutively as $\mathbf n_1,\ldots,\mathbf n_4$, the plaquette may be divided into two triangles in two distinct ways, corresponding to its two diagonals. Each triangle determines a spherical triangle on $S^2$, whose signed area $A$ is specified by the spins at its three vertices.

\begin{figure}
	\centering
	\includegraphics[width=.7\textwidth]{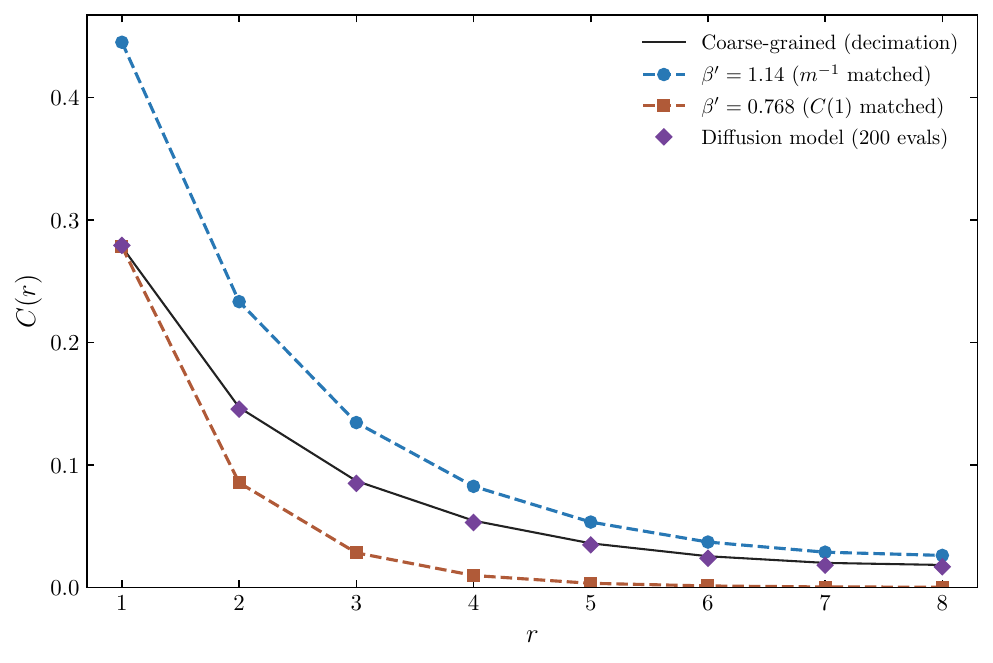}
	\caption{A two-point correlation function for an $O(3)$ model coarse-grained from a $64^2$ lattice ($\beta = 1.5$) to a $16^2$ lattice (black). One-parameter renormalizations of $\beta$ to $\beta'$, fitting either the correlation length (blue) or the energy per site (red), are compared to the generated configurations from a diffusion model (purple). See also Table~\ref{Table correlations16}.}\label{Fig correlations16}
\end{figure}

The two triangulations lead to two definitions of the local topological charge $q_i$
\begin{align}
	q^{(1)}_i = \frac{1}{4\pi}\left[
	A(\mathbf{n}_1,\mathbf{n}_2,\mathbf{n}_3)
	+A(\mathbf{n}_1,\mathbf{n}_3,\mathbf{n}_4)
	\right],\quad q^{(2)}_i = \frac{1}{4\pi}\left[
	A(\mathbf{n}_1,\mathbf{n}_2,\mathbf{n}_4)
	+A(\mathbf{n}_2,\mathbf{n}_3,\mathbf{n}_4)
	\right].
\end{align}
The total topological charge is then $Q^{(a)}=\sum_i q_i^{(a)}$.  An explicit formula for the signed area $A$ is given in \cite{bergLuscher1981}. The topological susceptibility is defined as
\begin{align}
	\chi_t= \frac{1}{V}\left\langle {Q^{(a)}}^2\right\rangle.
\end{align}

For a given plaquette $i$, the two conventions for local topological charge $q^{(a)}_i$ may either be strictly identical or they may differ by $\pm 1$. Such a disagreement signals a lattice-scale topological defect, which we will refer to as a \emph{dislocation} \cite{berg1981dislocations}.

Dislocations have long been recognized as an obstacle to determining the continuum topological susceptibility, and cooling procedures have been used to suppress them \cite{berg1981dislocations,diGiacomo1992}. However, semiclassical arguments \cite{luscher1982topology} and later studies \cite{bietenholzEtAl2018,yiThomasMonahan2022,bonanno2023} suggest that the dimensionless susceptibility $\chi_t m^{-2}$ diverges in the continuum limit due to contributions from arbitrarily small instantons. Here we restrict our attention to finite correlation lengths and use the distribution of $Q^{(a)}$ and the total number of dislocations as additional observables for assessing the accuracy of diffusion-generated configurations (see Fig.~\ref{Fig topCharge}).

\subsection{Numerical setup and baseline results}

The primary MCMC algorithm we use for generating a training set is the heat bath algorithm, in which each local spin $\mathbf{n}_i$ is drawn from a thermal distribution induced by its neighbor field $\mathbf{m}_i$,
\begin{align}
	p(\mathbf{n}_i | \mathbf{m}_i) \propto e^{\beta \mathbf{m}_i\cdot \mathbf{n}_i}. \label{Def vMF}
\end{align}
This is an example of the \emph{von Mises-Fisher} distribution on $S^2$ and it can be sampled efficiently. Spins are updated alternately on even and odd sublattices to form one heat bath sweep.

We generate a training set of $10^5$ configurations at $\beta=1.5$ on a periodic $64^2$ lattice. Starting from random configurations, we perform 1600 heat bath sweeps for thermalization before collecting the training configurations. Decimation is then used to produce coarse-grained training sets on the $32^2$ and $16^2$ lattices as well.

Distinct diffusion models are trained on these three resolutions of training set data. Detailed discussions of the sampling procedure (Sec.~\ref{Sec DM reverse diff}, \ref{Sec DM additional methods}), the training loss (Sec.~\ref{Sec DM heat kernel}, \ref{Sec Stein loss}), and the model architecture (Sec.~\ref{Sec DM score parameterization}) appear below. Here we present some basic results comparing model output to genuine equilibrium data obtained with the heat bath algorithm.

Unless otherwise stated, numerical results for the diffusion models are obtained from ensembles of $10^4$ generated configurations. These are distinct from the training ensemble of $10^5$ configurations and the reference ensemble of $0.9\times10^7$ configurations used to estimate equilibrium observables. The reference ensemble was generated in connection with the equilibration experiment discussed below (Fig.~\ref{Fig equilibration}), with each configuration obtained from an independent heat bath run.

In Fig.~\ref{Fig correlations16} and Table~\ref{Table correlations16}, the results of the diffusion model on the $16^2$ lattice ($\mathrm{DM}_{16}$) are compared with those obtained from $10^5$ configurations generated using conventional $O(3)$ models on a $16^2$ lattice, with effective couplings $\beta'$ tuned to reproduce either the correlation length or $C(1)$ of the coarse-grained reference ensemble. The diffusion model is sampled using both 1600 and 200 score evaluations, illustrating the tradeoff between computational cost and accuracy. Unlike the models with renormalized couplings, the diffusion model reproduces both short and long-distance correlations.

\begin{table}[t]
	\centering
	
	\begin{tabular}{lccc}
		\toprule
		Method & $C(1)$ & $m^{-1}$ & $\chi_m$ \\
		\midrule
		Equilibrium                    & $0.278183(6)$ & $2.6551(7)$ & $11.350(3)$ \\
		$\mathrm{DM}_{16}$ (1600 evals) & $0.2781(2)$   & $2.68(2)$   & $11.39(8)$  \\
		$\mathrm{DM}_{16}$ (200 evals)  & $0.2794(3)$   & $2.59(2)$   & $11.04(8)$  \\
		$\beta'=0.768$                 & $\emph{0.2782}$      & $1.06$      & $4.269$    \\
		$\beta'=1.14$                  & $0.4452$      & $\emph{2.66}$      & $16.41$    \\
		\bottomrule
	\end{tabular}
	\caption{Comparison of observables on the coarse-grained $16^2$ lattice
		with samples from both the diffusion model and $O(3)$ models
		at effective couplings $\beta'$. The observables used to determine each $\beta'$ are indicated in italics.}
	\label{Table correlations16}
	
\end{table}

\begin{figure}
	\centering
	\includegraphics[width=.8\textwidth]{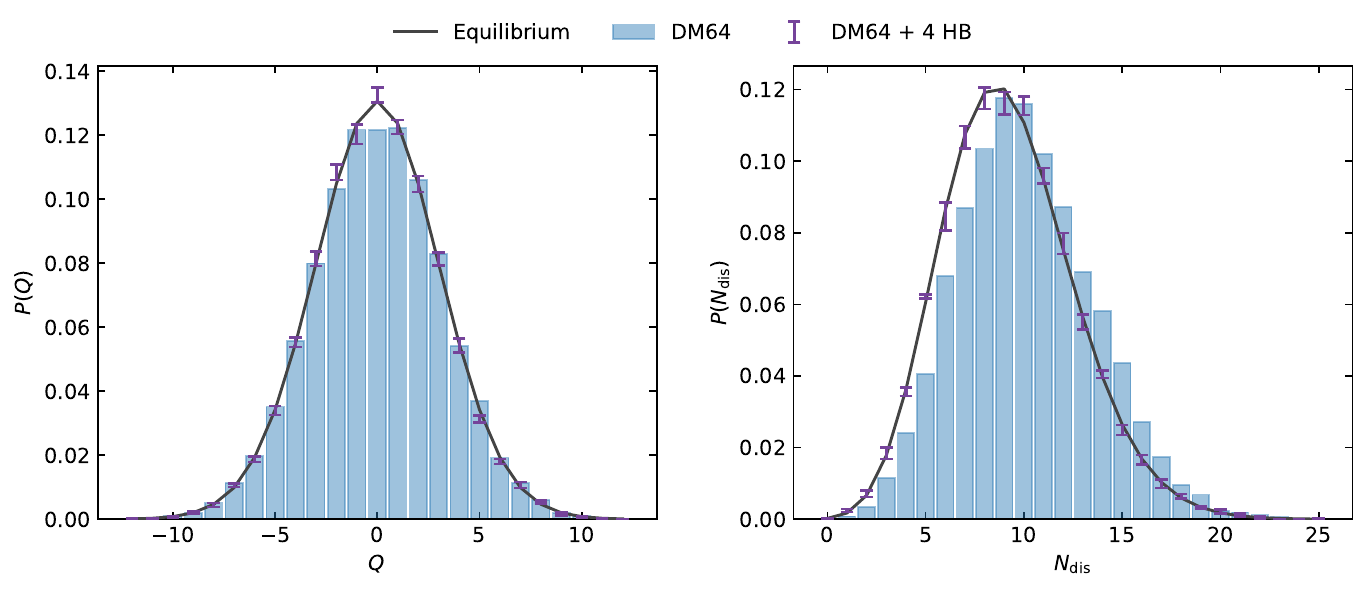}
	\caption{Distributions of the topological charge $Q$ (left) and the number of lattice dislocations $N_{\mathrm{dis}}$ (right) for the two-dimensional $O(3)$ model at $\beta=1.5$ on a $64^2$ lattice. The equilibrium heat-bath reference (black line) is compared with $10^4$ configurations generated by a diffusion model using 800 score evaluations (blue histogram). Systematic errors in the configurations generated by the model are largely corrected after a cheap refinement of four heat-bath sweeps (purple error bars).}\label{Fig topCharge}
\end{figure}

In Fig.~\ref{Fig topCharge}, the distributions of the topological charge $Q\equiv\left(Q^{(1)}+Q^{(2)}\right)/2$, and the number of dislocations are shown both for the reference ensemble and the output of $\mathrm{DM}_{64}$. The direct samples exhibit a systematic bias in the number of dislocations and a slightly reduced probability near $Q=0$. Both discrepancies are corrected after a few heat bath sweeps, suggesting that they arise from inaccuracies in the model at short distances. The correlation function $C(r)$ of $\mathrm{DM}_{64}$ is discussed later in connection with Fig.~\ref{Fig correlations64}. The values of $C(r)$ for direct samples from both $\mathrm{DM}_{64}$ and $\mathrm{DM}_{32}$ are compared to results from various super-resolution schemes in Table~\ref{Table superres}.

\begin{figure}
	\centering
	\includegraphics[width=.8\textwidth]{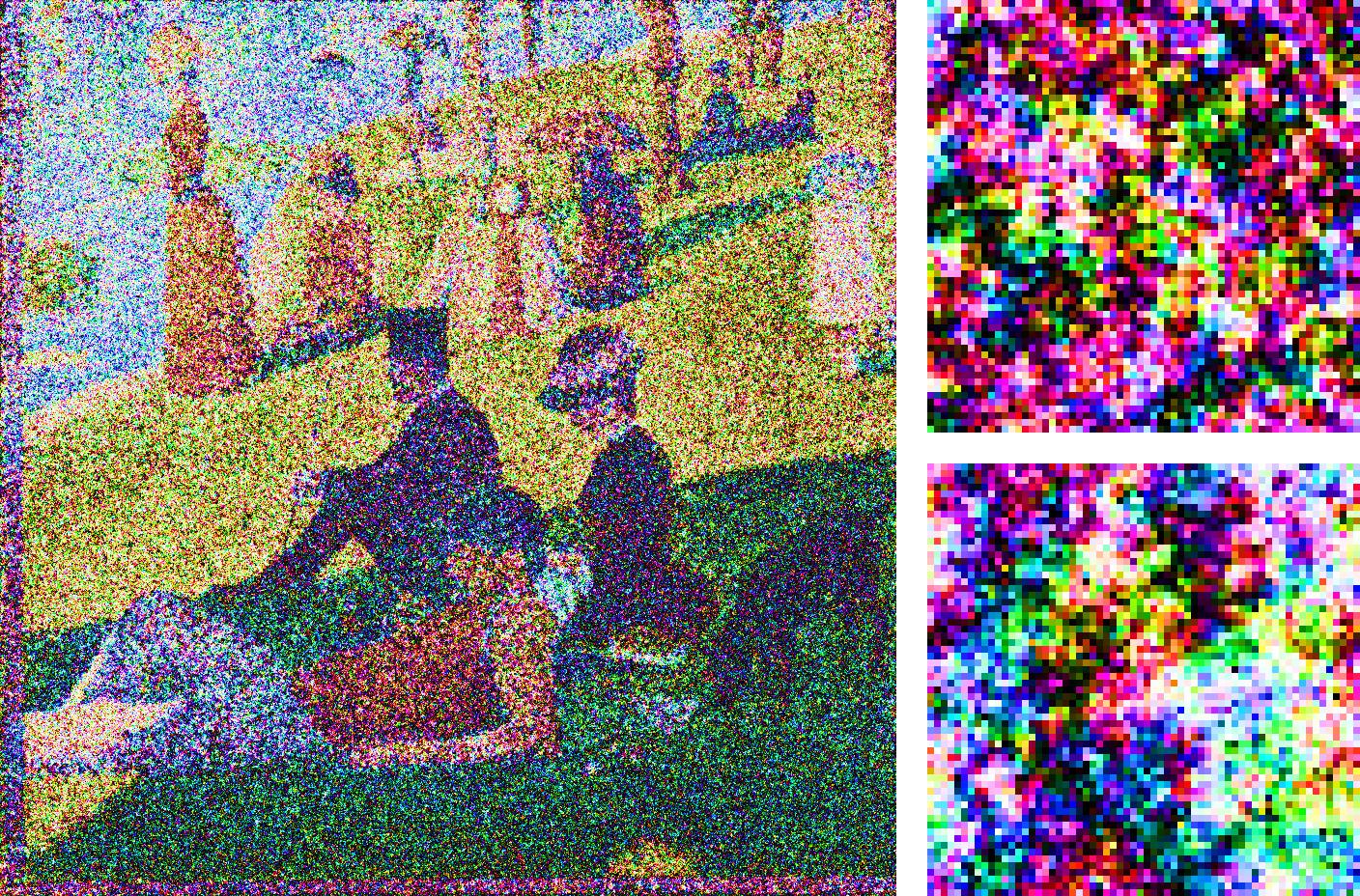}
	\caption{Editing short-distance structure with diffusion models. Configurations of classical spins are visualized by mapping spin orientations to saturated colors. An image of Seurat's painting \emph{A Sunday Afternoon on the Island of La Grande Jatte} is first converted to saturated colors and then to a corresponding spin configuration. A diffusion model trained on the $O(3)$ model at $\beta=1.5$ is then used to modify the short-distance structure in a manner consistent with the local statistics of the equilibrium field theory. At the upper right is an enlarged $64^2$ sample of the face of the man wearing a top hat. At the lower right is a genuine equilibrium $64^2$ configuration for comparison. The edited painting has an energy per site of $-1.201$ and a density of $2.40\times10^{-3}$ topological lattice dislocations, compared with $-1.203$ and $2.26\times10^{-3}$ in the equilibrium theory, respectively.}\label{Fig seurat}
\end{figure}

Figure~\ref{Fig seurat} illustrates both our method for visualizing lattice spin configurations and an application of the multiscale methods discussed later in Sec.~\ref{Sec superres}. In the HSL color representation, the space of fully saturated colors is topologically equivalent to $S^2$, allowing each spin orientation $\mathbf{n}$ to be associated with a unique color. This correspondence also allows an ordinary image to be converted into a lattice spin configuration, as illustrated using Seurat's painting. The short-distance structure of this configuration is then modified using $\mathrm{DM}_{64}$, following a procedure similar to the SDEdit algorithm \cite{meng2022}, which has also been applied to super-resolution of turbulent flows \cite{shuLiFarimani2023}. The resulting image has a value of $C(1)$ and density of lattice dislocations close to their equilibrium values, as reported in the figure caption. This procedure is closely related to method $B_1$ among the super-resolution schemes compared later in Fig.~\ref{Fig superres}.

\section{Diffusion models on $S^2$}
\label{Sec DM}

\subsection{Reverse diffusion}\label{Sec DM reverse diff}

The extension of continuous-time diffusion models \cite{song2021} to Riemannian manifolds largely follows by replacing the usual Euclidean constructions with their geometric counterparts. We briefly review the aspects relevant to our implementation, referring to \cite{deBortoli2022,huangEtAl2022} for a more detailed mathematical treatment.

The forward diffusion process transforms the target probability distribution at \(\tau=0\) into a distribution that is approximately uniform at the final diffusion time $\tau=\tau_f$,
\begin{align}
\der{	p}{\tau}=\nabla^2 p. \label{Eq forward diff}
\end{align}
Here $p$ is a density with respect to the Riemannian volume measure, and $\nabla^2$ is the Laplace-Beltrami operator, understood to act independently on the factors of $S^2$ at each lattice site.

Defining $\tilde{\tau} = \tau_f -\tau$, and using $\nabla^2 p = \nabla\cdot \left(p\nabla \log p\right)$, the forward equation may be rearranged to an equation in reverse time \cite{anderson1982}
\begin{align}
	\der{	p}{\tilde{\tau}}=-\nabla\cdot \left(2s\, p\right)+\nabla^2 p. \label{Eq reverse diff}
\end{align}
where the score $s = \nabla \log p$. This is a Fokker-Planck equation with a drift term determined by $s$, and as usual it is associated to a stochastic differential equation (SDE),
\begin{align}
d\mathbf{n}_i =  2 \mathbf{s}_i\, d\tilde{\tau} + \sqrt{2}\,d\mathbf{B}_i,
\end{align}
where $\mathbf{B}_i$ indicates Brownian motion on the factor of $S^2$ at lattice site $i$. Note that $\mathbf{s}_i$ implicitly depends on the field $\mathbf{n}$ at other lattice sites.

This SDE may be approximated for finite time steps $\Delta \tau$ in an Euler-Maruyama scheme \cite{deBortoli2022}
\begin{align}
	\mathbf{n}_i(\tau - \Delta \tau) = \text{Exp}_{\mathbf{n}_i(\tau)}\left[2 \mathbf{s}_i(\tau)\, \Delta \tau + \sqrt{2 \Delta \tau}\boldsymbol{\xi}_i\right]
\end{align}
where $\boldsymbol{\xi}_i$ is a random Gaussian vector of unit variance in the tangent space to $\mathbf{n}_i(\tau)$, and the exponential map for the sphere is concretely
\begin{align}
\text{Exp}_{\mathbf{n}}\left[\mathbf{v}\right]= \cos |\mathbf{v} |\,\, \mathbf{n} + \sin |\mathbf{v} |\, \frac{\mathbf{v}}{|\mathbf{v} |}, \qquad \mathbf{n}\cdot \mathbf{v}=0.
\end{align}

In practice we choose an alternate parametrization $u$ of the diffusion time $\tau$,
\begin{align}
	u(\tau) = \frac{\log\left(1 + \frac{\tau}{\tau_0}\right)}{\log\left(1 + \frac{\tau_f}{\tau_0}\right)}.
\end{align}
For large $\tau$ this is logarithmic in $\tau$ much like the $t$ parameter in \cite{song2021}, but here we have a regularization $\tau_0 = 0.005$, so that $u(0)=0$ is finite. We take the final diffusion time $\tau_f = 4$ and $u(\tau_f)=1$. The SDE is discretized in uniform steps in $u$, and the model is given $u$ as input. In training, the forward diffusion times are chosen uniformly in $u$, but the loss is reweighted to be uniform in $\tau$, as suggested in \cite{songEtAl2021likelihood}. For a broader discussion of implementation choices in diffusion models, see \cite{karras2022}.

\subsection{Additional reverse diffusion methods}\label{Sec DM additional methods}

The diffusion term in the forward diffusion equation \eqref{Eq forward diff} may also be rewritten entirely as drift,
\begin{align}
	\der{	p}{\tilde{\tau}}=-\nabla\cdot \left(s\, p\right). \label{Eq reverse diff ode}
\end{align}
This is the continuity equation associated with the ordinary differential equation (ODE) \cite{song2021},
\begin{align}
	\derOrd{\mathbf{n}_i}{\tilde{\tau}}= \mathbf{s}_i.\label{Eq ODE}
\end{align}
This is integrated with a second-order Runge-Kutta (RK2) method on $S^2$ \cite{leung2024}, which is essentially Heun's method evaluated with exponential maps. At fixed computational cost, the RK2 method requires approximately twice the step size of the Euler–Maruyama scheme, since it uses two score evaluations per step. In our numerical experiments, it nevertheless often performs marginally better when the learned score is sufficiently accurate.

We also make use of Langevin dynamics to correct the distribution at fixed diffusion time $\tau$ \cite{song2021},
\begin{align}
	d\mathbf{n}_i = \mathbf{s}_i(\tau)\,dv + \sqrt{2}\,d\mathbf{B}_i.\label{Eq langevin correction}
\end{align}
This SDE involves an auxiliary time $v$, independent of the diffusion time $\tau$, and has $p(\mathbf{n}; \tau)$ as its stationary distribution. It is integrated with finite steps $\Delta v$ using the same exponentially mapped Euler-Maruyama scheme as the reverse diffusion SDE.

\subsection{Score matching and the heat kernel}
\label{Sec DM heat kernel}

The score is approximated by a neural network $s_\theta(\mathbf{n}, \tau)$, trained using denoising score matching (DSM) \cite{vincent2011}. The loss is
\begin{align}
	\mathcal{L}_{\mathrm{DSM}}
	= \int_0^{\tau_f} d\tau\,
	\mathbb{E}_{\mathbf{n},\mathbf{n}'}
	\left[
	\sum_i
	\left|
	\mathbf{s}_{\theta,i}(\mathbf{n},\tau)
	- \nabla_{\mathbf{n}_i} \log p(\mathbf{n}_i|\mathbf{n}'_i; \tau)
	\right|^2
	\right].
	\label{Eq DSM loss}
\end{align}
The conditional distribution $p(\mathbf{n}_i|\mathbf{n}'_i;\tau)$ is the transition probability from an initial spin $\mathbf{n}'_i$ to the spin $\mathbf{n}_i$ at time $\tau$. This is given by the heat kernel on $S^2$, which is the propagator of the forward diffusion equation \eqref{Eq forward diff}.

Even on simple homogeneous manifolds such as $S^2$, the heat kernel does not have a closed-form expression, but can be represented as an infinite series. One alternative is to choose a simpler conditional distribution, such as the von Mises--Fisher distribution \eqref{Def vMF}, although this requires modifying the forward diffusion and reverse dynamics \cite{chemseddineEtAl2026}. Here we retain the conventional heat-kernel diffusion and discuss its numerical evaluation.

The spectral representation of the heat kernel is
\begin{align}
	p(\mathbf{n}_i|\mathbf{n}'_i;\tau)=\frac{1}{4\pi}\sum_{\ell=0}^\infty
	(2\ell+1)e^{-\ell(\ell+1)\tau}P_\ell(\cos\theta),
\end{align}
where $P_\ell$ are the Legendre polynomials, and $\mathbf{n}\cdot \mathbf{n}'=\cos \theta$. Due to the exponential factor this can be truncated reliably for large $\tau$. At small $\tau$, however, many oscillating terms must be summed, making direct evaluation inefficient and potentially numerically unstable.

For small $\tau$, we instead use the Minakshisundaram-Pleijel-Schwinger-DeWitt (MPSD) expansion which was developed further in the work of Camporesi \cite{camporesi1990}. The MPSD representation was emphasized in a diffusion model context already in \cite{lou2023}. The leading short-time approximation to the heat-kernel score on $S^2$ was also used in \cite{xiChen2026}. To extend the range of applicability to moderate values of $\tau$, we evaluate the expansion to higher order and develop a supplementary expansion to handle difficulties near $\theta=\pi$.

For low $\tau$, we take $p(\mathbf{n}_i|\mathbf{n}'_i;\tau)\approx K_{MPSD}(\theta, \tau)$ with
\begin{align}
	K_{MPSD}(\theta, \tau) 
	=
	\frac{e^{\tau/4}}{4\pi \tau}
	\sqrt{\frac{\theta}{\sin\theta}}
	\exp\left[-\frac{\theta^2}{4\tau}\right]F(\theta, \tau),\label{Eq MPSD}
\end{align}
\begin{align}
F(\theta,\tau)=1
+&\tau\left(
\frac{1}{12}
+\frac{\theta^2}{180}
+\frac{\theta^4}{1890}
+\frac{\theta^6}{18900}
\right)+\tau^2\left(
\frac{7}{480}
+\frac{13\theta^2}{5040}
+\frac{19\theta^4}{50400}
\right)\non
&\quad+\tau^3\left(
\frac{31}{8064}
+\frac{157\theta^2}{120960}
\right)
+\frac{127}{92160}\tau^4 + \order{\tau^5}.
\label{Eq MPSD correction}
\end{align}
Given the Gaussian factor in $K_{MPSD}$, we also expand $F(\theta,\tau)$ in $\theta$, using the power counting $\theta^2\sim \order{\tau}$. Appendix \ref{App MPSD} provides sufficient detail to extend the expansion of $F$ to arbitrary order, or to retain the full $\theta$ dependence of its coefficients.

The MPSD expansion is singular at $\theta=\pi$ due to the factor of $(\sin\theta)^{-1/2}$ in \eqref{Eq MPSD}, and becomes unreliable in a boundary layer $\pi-\theta\sim\order{\tau}$. This issue was noted in \cite{zhaoSong2018}. For small $\tau$, the probability of encountering this region is exponentially suppressed, but the problem becomes significant at moderate diffusion times.

In Appendix \ref{App BL} we develop a boundary-layer expansion for the heat kernel in this region, $p(\mathbf{n}_i|\mathbf{n}'_i;\tau)\approx K_{BL}(\eta,\tau)$, where the full dependence on the scaled variable
\begin{align}
	\eta=\frac{\pi(\pi-\theta)}{2\tau}\label{Def eta}
\end{align}
is retained at each order in $\tau$. We obtain
\begin{align}
	K_{BL}(\eta,\tau)
	&=
	\frac{\sqrt{\pi}e^{\tau/4}}{4\tau^{3/2}}
	e^{-\pi^2/(4\tau)}
	\left[
	I_0(\eta)
	-\frac{\tau}{\pi^2}
	\left(
	\eta^2 I_0(\eta)+\eta I_1(\eta)
	\right)
	+\order{\tau^2}
	\right].
	\label{Eq boundary layer}
\end{align}
At $\theta=\pi$, the quantity in brackets reduces to $1$, and we recover the leading antipodal expression appearing in \cite{hsu2002,barilariJendrej2014}, but I am unaware of any appearance of this boundary-layer expansion in the literature.

	\begin{figure}
			\centering
			\includegraphics[width=0.6\textwidth]{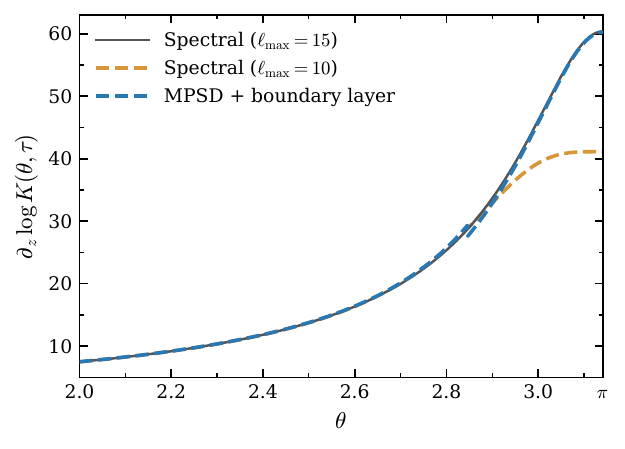}
			\caption{Comparison of approximations to the scalar score $\partial_z\log K(\theta,\tau)$, where $z=\cos\theta$. The spectral expansion with $\ell_{\max}=10$ is used at large $\tau$, while the MPSD and boundary-layer approximations are used at small $\tau$. These approximations are shown at the switching time $\tau=0.137$, alongside a higher-order spectral expansion ($\ell_{\max}=15$).}\label{Fig heat kernel score}
		\end{figure}

Given these approximations for the heat kernel, the conditional score appearing in the DSM loss \eqref{Eq DSM loss} may be calculated by straightforward differentiation. For the score, we switch between the spectral representation and the short-time approximations at $\tau=0.137$. Figure~\ref{Fig heat kernel score} compares the approximations at this switching time, including a higher-order spectral expansion as a reference. The switching time was chosen by requiring the heat-kernel-weighted relative mean squared score error to fall below a prescribed threshold.

In addition to evaluating the score, training requires sampling from the conditional distribution $p(\mathbf{n}_i|\mathbf{n}'_i;\tau)$ to perform the expectation in \eqref{Eq DSM loss}. We use rejection sampling with a von Mises--Fisher proposal distribution. For evaluation of the heat kernel itself, we use a separate switching time $\tau=0.15$, determined by comparing its integrated absolute error over $S^2$ against a higher-order spectral reference.

For larger $\tau$, we truncate the spectral expansion at $\ell_{\max}=10$ for the score and $\ell_{\max}=8$ for the heat kernel. We also developed a scheme for progressively reducing the number of terms as $\tau$ increases, but found this difficult to implement efficiently on a GPU. Details of this scheme, as well as the transition between the MPSD and boundary-layer approximations, are available in our publicly available code.

\subsection{Score parameterization and network architecture}\label{Sec DM score parameterization}

The learned score must be tangent to $S^2$, satisfying $\mathbf{n}_i\cdot\mathbf{s}_{\theta,i}=0$. One way to enforce this is to have the neural network output coefficients in a basis of tangent vector fields. However, the tangent bundle of $S^2$ is nontrivial, and no globally smooth basis exists.

As discussed in \cite{deBortoli2022}, this obstruction can be avoided using an overcomplete set of tangent vector fields. A convenient choice is the three rotational generators $\mathbf{e}^R_a=\hat{\mathbf{r}}_a\times\mathbf{n}$, where $\hat{\mathbf{r}}_a$ are the Cartesian unit vectors. The network then outputs three coefficients $\omega_\theta^a$, giving
\begin{align}
	\mathbf{s}_\theta
	=\sum_a\omega_\theta^a\mathbf{e}^R_a
	=\boldsymbol{\omega}_\theta\times\mathbf{n}.\label{Eq score overcomplete}
\end{align}
The component of $\boldsymbol{\omega}_\theta$ parallel to $\mathbf{n}$ is redundant and unconstrained by the loss. An equivalent parameterization, used in \cite{huangEtAl2022}, is to have the network output an unconstrained vector $\tilde{\mathbf{s}}_\theta$ and project it onto the tangent plane. This corresponds to choosing $\boldsymbol{\omega}_\theta=\mathbf{n}\times\tilde{\mathbf{s}}_\theta$.

An alternative is to provide the model with two orthonormal tangent vectors $\mathbf{e}^L_a$ alongside $\mathbf{n}$, allowing it to output the score as two components $s^a_\theta$ without redundancy. The choice of tangent frame introduces a $U(1)$ redundancy in the input. Together, $\mathbf{e}^L_a$ and $\mathbf{n}$ form the columns of an $SO(3)$ matrix, which can be represented by a unit quaternion $z\in S^3$ with an additional $\mathbb{Z}_2$ redundancy. Thus we lift the data from $\mathbf{n}\in S^2$ to $z\in S^3$, as in the construction underlying the $CP^1$ model \eqref{Def z}.

To further explore lifting the input to $S^3$, we train a model on an analytically known trimodal distribution on $S^2$, illustrated in Fig.~\ref{Fig smiley}. We find that the lifted network approximately learns the required gauge equivariance, although this property is not enforced by construction. In Appendix~\ref{App score parametrization}, we discuss these results in greater detail, as well as approaches to enforcing gauge equivariance and extensions to $CP^{N-1}$.

For the $O(3)$ lattice model, we adopt the overcomplete parameterization \eqref{Eq score overcomplete}. We use a convolutional U-Net based on standard score-based diffusion architectures \cite{ho2020,song2021,karras2022}, with three resolutions and channel widths $(16,32,64)$. The network initially uses two residual blocks at each resolution, with two additional blocks added at the finest resolution for the smaller lattice models. Diffusion time $u$ is incorporated through a learned scalar embedding added at each residual block. Global $O(3)$ symmetry is encouraged through random rotations of the training configurations.

\section{Stein and Schwinger-Dyson diagnostics}
\label{Sec Stein}

As seen in Table~\ref{Table correlations16}, the accuracy of observables estimated from model-generated samples depends on the number of score evaluations used during sampling. To assess the accuracy of the learned score independently of numerical sampling errors, we introduce Stein diagnostics, based on identities dating back to Stein \cite{stein1972}, see also \cite{xuMatsuda2020} for an application to distributions on $S^2$. These identities are closely related to the Schwinger-Dyson equations familiar from lattice field theory \cite{gottlob1996,shuklaEtAl2026}, and we begin with a derivation along those lines.

\subsection{The Stein residual}

Let $\mathcal C=(S^2)^V$ denote the configuration space, with local coordinates $\sigma^\mu$ and metric $g_{\mu\nu}$. The expectation of a scalar observable $O$ may be written in a form reminiscent of a path integral,
\begin{align}
	\langle O\rangle_p
	=\int_{\mathcal C}d\sigma\sqrt{g}\,O(\sigma)e^{\log p(\sigma)}.
\end{align}
A Schwinger--Dyson identity follows by applying the divergence theorem on $\mathcal C$,
\begin{align}
	0	&=\int_{\mathcal C}d\sigma\sqrt{g}\,\nabla_\mu\left(g^{\mu\nu}\partial_\nu O\,e^{\log p}\right)\nonumber\\
	&=\left\langle\nabla_\mu\left(g^{\mu\nu}\partial_\nu O\right)
	+
	g^{\mu\nu}\partial_\mu\log p\,\partial_\nu O
	\right\rangle_p.
\end{align}
The explicit metric factors show that the identity holds on curved configuration spaces. Recognizing the Laplace-Beltrami operator in the first term and the score $s_p=\nabla\log p$ in the second, we obtain
\begin{gather}
\left\langle\mathcal{D}\left[O; s_p\right]\right\rangle_p = 0,
	\label{Eq Stein stationary}\\
	\mathcal{D}\left[O; s_p\right] \equiv	 s_p\cdot\nabla O+\nabla^2 O.\label{Def D operator}
\end{gather}
The quantity $\left\langle\mathcal{D}\left[O; s_p\right]\right\rangle_p$ is referred to as the \emph{Stein residual}. The identity \eqref{Eq Stein stationary} involves the exact score $s_p$ associated with the probability distribution $p$ used in the expectation. In practice, we instead evaluate the residual using the learned score $s_\theta$, and a nonzero residual indicates a discrepancy between the learned and target scores that can be detected independently of the sampling method.

A non-zero Stein residual is closely connected with the evolution of an observable under the Langevin correction method which is frequently employed in diffusion model sampling. The Langevin SDE \eqref{Eq langevin correction} is associated to a Fokker-Planck equation for the probability $p(v)$ which is taken to be a function of the Langevin time $v$,
\begin{align}
	\frac{dp}{dv}= -\nabla\cdot\left(s_\theta \, p\right)+\nabla^2 p.
\end{align}
Multiplying by $O$ and integrating over $\mathcal{C}$ gives
\begin{align}
	\frac{d}{dv}\langle O\rangle_p = \left\langle\mathcal{D}\left[O; s_\theta\right]\right\rangle_p.\label{Eq Stein residual and Langevin}
\end{align}
So the Stein residual measures the rate at which Langevin correction changes the expectation of an observable.

Here $p$ is the distribution of configurations undergoing Langevin evolution. Of particular interest is evaluating the residual with the target distribution $p(\tau=0)$. If the learned score were exact, this distribution would remain unchanged under Langevin evolution, therefore a nonzero residual is a diagnostic for errors in the learned score, and particularly those errors which are liable to affect an observable of interest. In the next section, we illustrate these ideas by applying the Stein residual to the two-point correlation function $C(r)$.

\subsection{Stein diagnostic for the correlation function}
\label{Sec Stein diagnostics}

Equation \eqref{Def D operator} for $\mathcal{D}$ is written in  terms of general coordinates $\sigma$ on configuration space, but the $O(3)$ model has more structure and it may help to be clear. We write the configuration space coordinates as $\sigma^a_i$, where $i$ is a lattice site index (with no summation convention), and $a$ is a coordinate index on $S^2$. We write the three components of $\mathbf{n}_i$ explicitly as $n^A_i$.

We wish to choose $O =\sum_A n^A_i n^A_{i+j}$, so that $\langle O\rangle = C(j)$. First consider the operator  $\mathcal{D}$ applied to $n^A_i$, with $\nabla_j^2$ referring to the Laplace-Beltrami operator acting on site $j$,
\begin{align}
	\mathcal{D}n^A_i &= \sum_j \left(s^a_j \partial_{\sigma^a_j} + \nabla^2_j \right)n^A_i \non
	&=  s^a_i \partial_{\sigma^a_i} n^A_i + \nabla^2_i n^A_i \non
	&=  s^A_i - 2 n^A_i. \label{Eq stein corr lemma}
\end{align}
In the first term of the last equality, $s^a_i \left(\partial_{\sigma^a_i} n^A_i\right) = s_i^A $ may be understood as a transformation between $a$ and $A$ components. The second term in the equality makes use of the spherical geometry. The components $n^A$ are $\ell=1$ spherical harmonics as functions of $\sigma^a$, and are eigenfunctions of the Laplace-Beltrami operator
\begin{align}
	\nabla^2 n^A = -\ell(\ell+1)n^A = -2 n^A.
\end{align}

Now restoring the boldface notation, $O =\mathbf{n}_i\cdot \mathbf{n}_{i+j}$ with $j\neq 0$,
\begin{align}
	\mathcal{D}O = \mathbf{s}_i\cdot \mathbf{n}_{i+j}+\mathbf{n}_i\cdot \mathbf{s}_{i+j}-4\mathbf{n}_i\cdot \mathbf{n}_{i+j},
\end{align}
so the Stein residual for $O$ is
\begin{align}
	\left\langle\mathcal{D}O\right\rangle = \left\langle\mathbf{s}_i\cdot \mathbf{n}_{i+j}+\mathbf{n}_i\cdot \mathbf{s}_{i+j} \right\rangle - 4 C(j). \label{Eq stein corr}
\end{align}

In practice we use this by defining the Stein diagnostic for the correlation function $C_{\text{Stein}}$,
\begin{align}
	C_{\text{Stein}}(j)\equiv \frac{1}{4}\left\langle\mathbf{s}_{\theta,i}\cdot \mathbf{n}_{i+j}+\mathbf{n}_i\cdot \mathbf{s}_{\theta,i+j} \right\rangle,\label{Def Stein diagnostic}
\end{align}
where the expectation value is taken over the training set.

\begin{figure}
	\centering
	\includegraphics[width=0.8\textwidth]{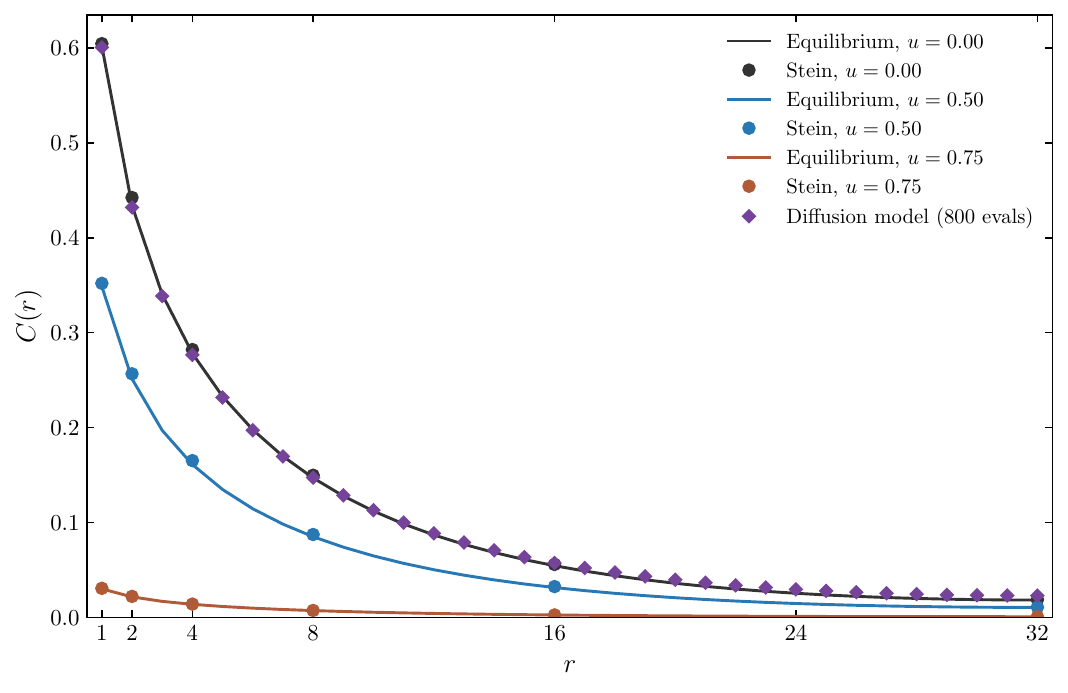}
	\caption{Two-point correlation function $C(r)$ on the $64^2$ lattice. Solid lines show equilibrium correlations at diffusion times $u=0$, $0.5$, and $0.75$, while circles show the corresponding Stein estimates \eqref{Def Stein diagnostic}. Diamonds show correlations measured from configurations generated by the diffusion model using 800 score evaluations.}\label{Fig correlations64}
\end{figure}

$C_{\text{Stein}}(j)=C(j)$ if the learned score is equal to the target score, so it is tempting to consider $C_{\text{Stein}}$ as an estimate of the correlation function calculated directly from the model score. No sampling with reverse diffusion is required. Indeed consider the results in Fig.~\ref{Fig correlations64}. The Stein diagnostic calculated with a well-trained model closely tracks the equilibrium correlation function over a wide range of diffusion times including $u=0$.

The residual $4\left(C_{\text{Stein}}(j)-C(j)\right)$ also provides qualitative information about the errors we might expect for correlation functions calculated from model-generated samples. From Eq.~\eqref{Eq stein corr} and \eqref{Eq Stein residual and Langevin}, a positive residual indicates that the learned score would instantaneously increase the value of $C(j)$ under Langevin evolution, and thus one might expect the stationary state under Langevin correction to have an overcorrelated value of $C(j)$. Conversely, a negative residual would suggest undercorrelation. This is not guaranteed, since the Stein residual describes the instantaneous evolution of the correlation function rather than its value in the stationary distribution.

Interestingly, we find that models producing the most accurate correlations during sampling often have slightly overcorrelated Stein diagnostics (consider Table~\ref{Table Stein correlations}). One possible explanation is a cancellation of systematic errors. Sampling with a finite number of score evaluations tends to produce undercorrelated configurations, but this may be compensated by the errors in the learned score which favor stronger correlations.

\begin{table}
	\centering
		\caption{Comparison of correlations in the equilibrium reference ensemble, samples from the diffusion models, and Stein diagnostic estimates of the correlation function. On the $64^2$ lattice, three models were used. Model $A$ was determined as the best checkpoint for sampling and Models $B$ and $C$ were trained with and without a $C(1)$ Stein residual loss term.}
	\label{Table Stein correlations}
	\begin{tabular}{cc l cccc}
		\toprule
		Lattice & $u$ & Method
		& $C(1)$ & $C(2)$ & $C(4)$ & $C(8)$ \\
		\midrule
		$64^2$ & $0$
		& Equilibrium
		& 0.601683(2) & 0.433864(4) & 0.278183(6) & 0.147033(8) \\
		& & $\mathrm{DM}_{A}$ (800 evals)
		& 0.60031(8) & 0.43188(9) & 0.2767(1) & 0.1473(2) \\
		& & $\mathrm{DM}_{A}$ (200 evals)
		& 0.59803(7) & 0.4269(1) & 0.2700(1) & 0.1405(1) \\
		& & $\text{Stein}_A$
		& 0.60429 & 0.44230 & 0.28220 & 0.14989 \\
		& & $\text{Stein}_{B}$
		& 0.60087 & 0.44004 & 0.28040 & 0.14901 \\
		& & $\text{Stein}_{C}$
		& 0.59751 & 0.43700 & 0.27888 & 0.14815 \\
		\addlinespace
		& $0.5$ & Equilibrium
		& 0.348519(1) & 0.251311(2) & 0.161135(3) & 0.085167(5) \\
		& & $\text{Stein}_{A}$
		& 0.35192 & 0.25666 & 0.16528 & 0.08748 \\
		& & $\text{Stein}_{B}$
		& 0.34870 & 0.25421 & 0.16387 & 0.08675 \\
		& & $\text{Stein}_{C}$
		& 0.34617 & 0.25210 & 0.16233 & 0.08594  \\
		\specialrule{1.2pt}{8pt}{5pt}
		$16^2$ & $0$ & Equilibrium
		& 0.278183(6) & 0.147033(8) & 0.05464(1) & 0.01848(2) \\
		& & DM (1600 evals)
		& 0.2781(2) & 0.1467(3) & 0.0545(3) & 0.0191(5) \\
		& & DM (200 evals)
		& 0.2794(3) & 0.1457(4) & 0.0531(4) & 0.0170(4) \\
		& & Stein
		& 0.27975 & 0.14735 & 0.0544 & 0.0183 \\
		\bottomrule
	\end{tabular}

\end{table}

\subsection{Training with Stein loss}\label{Sec Stein loss}

The Stein residual can also be used to construct an additional training loss, with the aim of penalizing errors in the learned score that affect particular observables. A related approach using loss functions constructed from Schwinger-Dyson identities has been developed in \cite{shuklaEtAl2026}. In principle, we may introduce a loss of the form
\begin{align}
	\mathcal L_{\text{Stein}}
	=\int_0^1 du\,\left(\left\langle\mathcal D \left[O; s_\theta(u)\right]\right\rangle_{p(u)}\right)^2,
\end{align}
The expectation value is taken before squaring, so that the loss vanishes when the Stein identity is satisfied.

In practice, since the diffusion times are chosen randomly for each training sample, it is difficult to estimate the expectation over $p(u)$ at a fixed time. We divide the diffusion time interval into $K$ equal bins $I_k$ and average over configurations and times within each bin before squaring,
\begin{align}
	\mathcal L_{\text{Stein}}
	=\frac{1}{K}\sum_{k=1}^K\,\left(K\int_{I_k}du\left\langle\mathcal D \left[O; s_\theta(u)\right]\right\rangle_{p(u)}\right)^2.
\end{align}
This loss is added to the usual denoising score matching loss,
\begin{align}
	\mathcal L=\mathcal L_{\text{DSM}}+\lambda\mathcal L_{\text{Stein}},
\end{align}
We use $K=4, \lambda=10$ in the experiments reported below.

We tested this approach using the Stein residual for $C(1)$, with the aim of improving the short-distance correlations of model-generated samples. For both $\mathrm{DM}_{64}$ and $\mathrm{DM}_{32}$, we continued the training of an already well-trained model checkpoint using the additional Stein loss, and compared it to a control model trained from the same checkpoint using only the DSM loss.

The Stein-trained models generally produce more accurate Stein diagnostics than the controls, including at larger separations, even though the additional loss involves only $C(1)$. This is seen in Table~\ref{Table Stein correlations}, where the Stein-trained model and the control model are labeled B and C, respectively. In the $64^2$ case, the additional training did not improve sampling accuracy beyond that of our original best checkpoint, labeled A. In the $32^2$ case, we selected the Stein-trained model for the super-resolution experiments presented below.

\section{Super-resolution methods}
\label{Sec superres}

\begin{table}
	\centering
	
	\begin{tabular}{llcc}
		\toprule
		Hardware & Method & it/s & $\mathrm{DM}_{64}$ evals \\
		\midrule
		\multirow{4}{*}{RTX PRO 6000 Blackwell}
		& $\mathrm{HB}_{64}$ & 31.03 & 0.0393 \\
		& $\mathrm{DM}_{16}$ & 12.16 & 0.100 \\
		& $\mathrm{DM}_{32}$ & 3.19  & 0.382 \\
		& $\mathrm{DM}_{64}$ & 1.22  & 1.000 \\
		\midrule
		\multirow{4}{*}{A100-SXM4-40GB}
		& $\mathrm{HB}_{64}$ & 23.8 & 0.0580 \\
		& $\mathrm{DM}_{16}$ & 11.3 & 0.122 \\
		& $\mathrm{DM}_{32}$ & 3.50 & 0.394 \\
		& $\mathrm{DM}_{64}$ & 1.38 & 1.000 \\
		\bottomrule
	\end{tabular}
	\caption{Computational costs of heat bath sweeps and diffusion model
		score evaluations on two GPUs. The final column gives the cost of
		each operation in equivalent $\mathrm{DM}_{64}$ score evaluations,
		calculated from the measured iteration rates.}
	\label{Table computational cost}
\end{table}

In the preceding sections, we have demonstrated that diffusion models can learn the distribution of the $O(3)$ model with reasonable accuracy. However, the computational cost of evaluating the learned score remains a significant obstacle to their use in lattice field theory. Table~\ref{Table computational cost} compares the cost of heat-bath sweeps and diffusion model score evaluations on two GPUs. Although the precise ratios depend on the hardware and neural network architecture, a full heat-bath sweep ($\text{HB}_{64}$) costs less than $6\%$ of a single score evaluation of $\mathrm{DM}_{64}$. Consequently, generating a configuration using hundreds of score evaluations may be more expensive than equilibrating a random configuration using the heat-bath algorithm, even though the diffusion-generated configuration may retain significant systematic errors.

The computational cost can be reduced considerably by sampling on smaller lattices. For example, a score evaluation of $\mathrm{DM}_{16}$ is only about twice as expensive as a heat-bath sweep on the $64^2$ lattice. This motivates a multiscale approach in which configurations are generated using inexpensive coarse-lattice diffusion models and super-resolved to finer lattices. We investigate whether such schemes can produce more accurate configurations at lower computational cost than direct sampling with $\mathrm{DM}_{64}$ alone. We also investigate the use of super-resolved configurations to accelerate equilibration under conventional heat-bath Monte Carlo in Appendix~\ref{App equilibration}.

Machine learning methods for super-resolution have previously been explored in the context of lattice field theory \cite{efthymiouBeachMelko2019} and there have been a number of applications involving normalizing flows \cite{shiina2024,bauer2024,singha2026}. At a high level, our approach begins with ordinary reverse diffusion sampling using $\mathrm{DM}_{16}$, followed by \emph{prolongation} to a finer lattice. This involves placing the coarse spins on a sublattice of the finer grid and interpolating the missing spins. Unlike approaches such as \cite{efthymiouBeachMelko2019}, our interpolation procedure itself does not involve machine learning. Instead, the interpolated configuration is refined in a separate step, using either conventional MCMC or the learned score on the finer lattice.

\begin{figure}
	\centering
	\includegraphics[width=\textwidth]{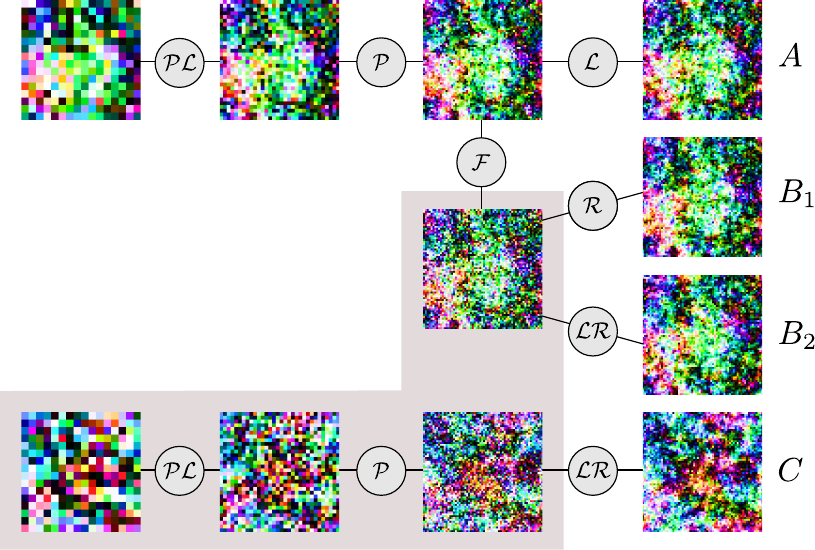}
	\caption{Illustration of various super-resolution schemes. The operator $\mathcal P$ denotes prolongation to a lattice with twice the linear size, while $\mathcal L$ denotes Langevin correction with the coarse spins held fixed. The operators $\mathcal F$ and $\mathcal R$ denote forward and reverse diffusion, respectively. The shaded region corresponds to diffusion time $u=u_{\mathrm{noise}}=0.5$, while the unshaded region corresponds to $u=0$. In method A, all prolongation and refinement steps are performed at $u=0$, following the initial generation of a $16^2$ configuration by reverse diffusion. In methods $B_1$ and $B_2$, the configuration is first prolonged to $64^2$ at $u=0$, then forward diffused to $u_{\mathrm{noise}}$. Method $B_1$ uses the remaining score evaluations entirely for reverse diffusion to $u=0$, whereas $B_2$ divides them between Langevin correction at $u_{\mathrm{noise}}$ and reverse diffusion. In method $C$, the initial $16^2$ configuration is reverse diffused only to $u_{\mathrm{noise}}$, where prolongation and refinement are performed. The final stages then follow method $B_2$.}\label{Fig superres}
\end{figure}

Our approach is closely related to multiscale MCMC methods that predate machine learning. In particular, Brandt and Ron \cite{brandtRon2001} considered conditional refinement in which the coarse degrees of freedom are held fixed while the finer degrees of freedom are updated, an approach we also employ. Endres et al.~\cite{endres2015} investigated the use of prolongated coarse configurations to reduce equilibration times in lattice gauge theory. An important difference is that our coarse-grained distributions are not described by a simple renormalized nearest-neighbor action, as illustrated in Fig.~\ref{Fig correlations16}. Constructing an effective action suitable for conventional MCMC is therefore nontrivial (see e.g. \cite{kushnirRosenstein1995extended}), whereas a diffusion model can learn the score of the coarse-grained distribution directly from training configurations.

We investigate several super-resolution schemes, illustrated in Fig.~\ref{Fig superres}. In each case, configurations are generated using $\mathrm{DM}_{16}$ with RK2 sampling \eqref{Eq ODE} and then progressively prolonged to finer lattices. Prolongation is implemented using conditional Gaussian interpolation \cite{hoffmanRibak1991}, which preserves the coarse spins and introduces new unnormalized spins according to a Gaussian distribution with a correlation function chosen to match that of the training set. It is discussed further in Appendix \ref{App Interp}.

The interpolated configurations are then normalized and refined using the learned score through Langevin correction with the coarse spins held fixed. In addition to refinement at $u=0$ ($A$), we consider schemes ($B_1,B_2$) in which configurations are forward diffused to $u_{\mathrm{noise}}=0.5$ before undergoing reverse diffusion. We also consider a related scheme ($C$) in which all prolongation takes place at $u_{\mathrm{noise}}$.

The four schemes also generate configurations on the intermediate $32^2$ lattice, as illustrated in the second column of Fig.~\ref{Fig superres}. There are two distinct procedures at this resolution, one shared by $A,B_1,B_2$ at $u=0$, and the other corresponding to $C$ at $u=u_{\mathrm{noise}}$.

Table~\ref{Table superres} compares the resulting correlation functions with those obtained from direct diffusion model sampling. Computational costs are expressed in equivalent score evaluations of the model on the target lattice, using the A100 timings in Table~\ref{Table computational cost}. For comparison, we also report the correlations obtained from Gaussian interpolation without subsequent refinement using the learned score.

For $\mathrm{DM}_{32}$, both super-resolution methods produce correlations substantially closer to equilibrium than direct diffusion sampling, at slightly lower computational cost. For $\mathrm{DM}_{64}$, the results depend more strongly on the reconstruction scheme. Method $B_2$ gives the best overall agreement with the reference correlations, improving upon direct sampling at comparable computational cost. In particular, we obtain $C(1)=0.60125(6)$, compared with $0.60031(8)$ from direct sampling and the reference value $0.601683(2)$. The comparison of $B_2$ with $B_1$, which closely resembles SDEdit \cite{meng2022}, may illustrate the importance of Langevin correction with fixed coarse spins at the intermediate diffusion time. Overall, these results demonstrate that multiscale diffusion sampling can improve accuracy without increasing computational cost.

\section{Discussion}
\label{Sec discussion}

\begin{table*}
	\centering
	
	\begin{tabular}{lcccccc}
		\toprule
		Method & $C(1)$ & $C(2)$ & $C(4)$ & $C(8)$ & $C(16)$
		& Evals \\
		\midrule
		\multicolumn{7}{l}{$\boldsymbol{32^2}$ lattice, $\boldsymbol{u=0}$} \\
		\midrule
		\textbf{Ref.} & 0.433864(4) & 0.278183(6) & 0.147033(8) & 0.05464(1) & 0.01848(2) & --- \\
		AB            & 0.4316(2) & 0.2772(2) & 0.1450(2) & 0.0530(3) & 0.0172(5) & 724 \\
		Direct        & 0.4233(1) & 0.2690(2) & 0.1398(2) & 0.0508(3) & 0.0165(3) & 800 \\
		Interp.       & 0.4035(3) & 0.2590(3) & 0.1368(3) & 0.0495(4) & 0.0160(4) & 124 \\
		\midrule
		\multicolumn{7}{l}{$\boldsymbol{32^2}$ lattice, $\boldsymbol{u=0.5}$} \\
		\midrule
		\textbf{Ref.} & 0.251311(2) & 0.161135(3) & 0.085167(5) & 0.031648(7) & 0.010704(9) & --- \\
		C             & 0.2482(1) & 0.1594(2) & 0.0831(2) & 0.0306(2) & 0.0101(3) & 722 \\
		Direct        & 0.2463(1) & 0.1562(1) & 0.0819(2) & 0.0300(2) & 0.0097(2) & 800 \\
		Interp.       & 0.2264(3) & 0.1454(3) & 0.0769(2) & 0.0287(3) & 0.0096(4) & 62 \\
		\midrule
		\multicolumn{7}{l}{$\boldsymbol{64^2}$ lattice, $\boldsymbol{u=0}$} \\
		\midrule
		\textbf{Ref.} & 0.601683(2) & 0.433864(4) & 0.278183(6) & 0.147033(8) & 0.05464(1) & --- \\
		$B_2$         & 0.60125(6) & 0.4333(1) & 0.2772(2) & 0.1452(3) & 0.0527(4) & 785 \\
		Direct        & 0.60031(8) & 0.43188(9) & 0.2767(1) & 0.1473(2) & 0.0575(4) & 800 \\
		A             & 0.59962(8) & 0.4304(1) & 0.2737(2) & 0.1437(2) & 0.0524(3) & 785 \\
		C             & 0.59818(9) & 0.4281(1) & 0.2713(2) & 0.1414(3) & 0.0518(4) & 784 \\
		$B_1$         & 0.59367(6) & 0.4253(1) & 0.2724(2) & 0.1431(2) & 0.0520(4) & 785 \\
		Interp.       & 0.58272(9) & 0.4220(1) & 0.2722(2) & 0.1428(2) & 0.0521(3) & 285 \\
		\bottomrule
	\end{tabular}
	\caption{Comparison of $C(r)$ obtained from the reference ensemble (Ref.), direct
		diffusion model sampling, and various super-resolution methods (see Fig.~\ref{Fig superres}) including Gaussian interpolation without final DM correction (Interp.).
		Computational costs are expressed in equivalent score evaluations (Evals)
		of $\mathrm{DM}_{32}$ for the $32^2$ lattice and $\mathrm{DM}_{64}$
		for the $64^2$ lattice, using the A100 timings reported in
		Table~\ref{Table computational cost}.}
	\label{Table superres}
\end{table*}

We have demonstrated that score-based diffusion models can reproduce the equilibrium distribution of the two-dimensional $O(3)$ model with reasonable accuracy. Direct sampling reproduces the two-point correlation functions over a range of distances (Figs.~\ref{Fig correlations16} and \ref{Fig correlations64}), as well as the distributions of topological charge and lattice dislocations (Fig.~\ref{Fig topCharge}), with further improvements obtained through relatively inexpensive Monte Carlo refinement.

Implementing diffusion models on the spherical target space also required addressing several technical challenges that are relevant beyond the $O(3)$ model. In particular, we developed a boundary-layer approximation to the heat kernel on $S^2$ in Sec.~\ref{Sec DM heat kernel}, which extends the range of diffusion times over which the efficient short-time representation can be used accurately. This approximation may also be useful for other applications of diffusion models on spherical manifolds.

Although we have demonstrated that diffusion models can reproduce the equilibrium distribution of the $O(3)$ model with reasonable accuracy, their practical use in lattice field theory requires confronting the computational cost of evaluating the learned score. To limit this cost, it is necessary to perform numerical integration using relatively few score evaluations, making it difficult to disentangle errors in the learned score from errors introduced by the sampling procedure.

In Sec.~\ref{Sec Stein diagnostics}, we introduce Stein diagnostics that allow us to investigate the accuracy of the learned score independently of reverse diffusion sampling. Loosely speaking, these diagnostics provide estimates of how well the correlation functions have been learned by the model at any diffusion time. Our particular construction \eqref{Def Stein diagnostic} makes essential use of the spherical geometry, although similar diagnostics could be constructed for correlations of Laplace-Beltrami eigenfunctions on other homogeneous spaces and Lie groups. More generally, Stein residuals measure the rate of change of observables under  Langevin correction, a procedure commonly employed to improve the accuracy of diffusion models.

Our other methodological focus was also motivated by the computational cost of score evaluations. In Sec.~\ref{Sec superres}, we investigated multiscale sampling methods that reduce this cost by performing part of the generation process on smaller lattices. An attractive application is to theories for which an explicit action suitable for conventional MCMC is unavailable or difficult to construct. This can arise in nonequilibrium systems or in coarse-grained field theories, where integrating out short-distance degrees of freedom produces a complicated effective action. Diffusion models offer the possibility of learning these distributions directly from training configurations, without requiring an explicit expression for the effective action. We demonstrated that the super-resolution schemes illustrated in Fig.~\ref{Fig superres} can simultaneously improve the accuracy of sampled correlation functions and modestly reduce computational cost compared with direct diffusion sampling on the finest lattice.

There remains considerable room for improving these methods. In particular, the computational cost is dominated by refinement on the finest lattice. More efficient numerical integration and improvements to the interpolation procedure could reduce the amount of fine-lattice refinement required. The latter may be particularly promising for scalar field theories, where conditional Gaussian interpolation can reproduce the two-point correlation function without requiring normalization onto a nonlinear target manifold, potentially reducing the subsequent refinement cost. Another possibility is to train diffusion models specifically for conditional reconstruction, rather than using models trained for unrestricted sampling to refine configurations with the coarse spins held fixed.

A further direction is to combine multiscale sampling with diffusion processes that introduce spatial structure progressively across different length scales. For example, the SKILD method \cite{chen2026skild} uses frequency-dependent noising in Fourier space, so that long-wavelength modes are reconstructed before short-wavelength modes during reverse diffusion (see also \cite{cotlerRezchikov2023}). This suggests a natural extension of our multiscale approach. Rather than performing reverse diffusion at a fixed lattice resolution, one could begin with a coarse model and progressively switch to models at higher resolutions as shorter-wavelength modes become relevant. Such a construction would require coordinating the diffusion processes across resolutions, but could substantially reduce computational cost by reserving expensive fine-lattice score evaluations for the final stages of generation.

\section*{Code availability}

The code associated with this work, including trained diffusion models and Jupyter notebooks illustrating the numerical methods, is publicly available at \url{https://github.com/daschubring/o3-cp1-diffusion}.

\appendix

\section{Short-time heat kernel asymptotics on $S^2$}

In Sec.~4.3 of \cite{camporesi1990}, Camporesi derives the heat kernel on $S^2$ by integrating the heat kernel on $SO(3)$ over the $SO(2)$ fiber, using results from harmonic analysis on compact Lie groups. A summary of the relevant points of the argument is given here.

The heat kernel on $SO(3)$ in Camporesi's normalization\footnote{In this normalization $SU(2)$ is equivalent to the three-sphere with radius $2$.} is
\begin{equation}
	K_{SO(3)}(\phi,\tau)
	=
	\frac{e^{\tau/4}}
	{2(4\pi\tau)^{3/2}}
	\sum_{n\in\mathbb Z}
	(-1)^n
	\frac{\phi+2\pi n}{\sin(\phi/2)}
	\exp\left[
	-\frac{(\phi+2\pi n)^2}{4\tau}
	\right].
	\label{Eq hk SO3}
\end{equation}
An approximation to this exact representation,
truncated to the three terms $n=0,\pm1$,
has also been employed in diffusion models on $SO(3)$
\cite{jagvaral2024}.

$S^2$ can be understood as the homogeneous space $SO(3)/SO(2)$. Its heat kernel is obtained by integrating the $SO(3)$ heat kernel over the isotropy subgroup $SO(2)$. Concretely,
\begin{equation}
	K_{S^2}(\theta,\tau)
	=
	\int d\alpha \,
	K_{SO(3)}(R_y(\theta)R_z(\alpha),\tau),
	\label{Eq hk S2 fiber}
\end{equation}
where $R_y(\theta)$ is a representative rotation taking the north pole of $S^2$
to a point at angular distance $\theta$, and $R_z(\alpha)$ is a rotation that leaves the north pole fixed.

To apply \eqref{Eq hk SO3} in \eqref{Eq hk S2 fiber}, we must calculate the equivalent rotation angle $\phi$ of $R_y(\theta)R_z(\alpha)$. For any rotation matrix $R$, $\text{Tr}\,R = 1+2 \cos \phi,$ thus
\begin{align*}
	\text{Tr}\left(R_y(\theta)R_z(\alpha)\right)=\cos\theta + \cos\alpha + \cos\theta\cos\alpha = 1+2\cos\phi.
\end{align*} 
After some manipulation,
\begin{gather}
	\cos^2 (\phi/2) = \cos^2 (\theta/2)\cos^2 (\alpha/2),\\
	\frac{\cos\theta -\cos\phi}{2} = \cos^2 (\theta/2)\sin^2 (\alpha/2).
\end{gather}
These identities are applied to express \eqref{Eq hk S2 fiber} as an integral over $\phi$,
\begin{equation}
	K_{S^2}(\theta,\tau)
	=
	\frac{\sqrt{2}\,e^{\tau/4}}{(4\pi\tau)^{3/2}}
	\sum_{n\in\mathbb Z}(-1)^n
	\int_\theta^\pi d\phi\,
	\frac{
		(\phi+2\pi n)
		\exp[-(\phi+2\pi n)^2/(4\tau)]
	}{
		\sqrt{\cos\theta-\cos\phi}
	}.
	\label{Eq hk S2 exact}
\end{equation}

\subsection{MPSD expansion}\label{App MPSD}

For fixed $0<\theta<\pi$ and for small $\tau$, we keep only the $n=0$ term in \eqref{Eq hk S2 exact}, and introduce variable $u=\phi-\theta$,
\begin{gather}
	K_{S^2}(\theta,\tau)\approx \frac{e^{\tau/4}}{4\pi \tau}
	\sqrt{\frac{\theta}{\sin\theta}}
	\exp\left(-\frac{\theta^2}{4\tau}\right)F(\theta, \tau),\non
	F(\theta,\tau)=\frac{\sqrt{\sin\theta}}{\sqrt{2\pi\tau\theta}}\int_0^{\pi-\theta}du\,
	\left[\frac{(\theta+u)e^{-u^2/(4\tau)}}{\sqrt{\cos\theta-\cos(\theta+u)}}\right]	e^{-\frac{\theta}{2\tau}u}.
\end{gather}
This is now in the MPSD form \eqref{Eq MPSD}, with the integral in $F$ dominated by the endpoint at $u=0$, with no interior stationary point. We introduce $v = \theta u/(2\tau)$ and expand the bracketed expression in the integrand in powers of $\tau$. The upper limit of the integral is taken to infinity, discarding terms which are exponentially small at fixed $\theta$,
\begin{align}	F(\theta,\tau)&\approx\frac{1}{\sqrt{\pi}}\int_0^{\infty}\frac{dv}{\sqrt{v}}\,
	\left[1+\left(\left(\frac{2}{\theta^2}-\frac{\cot\theta}{2\theta}\right)v-\frac{v^2}{\theta^2}\right)\tau + \order{\tau^2}\right]	e^{-v}\non
	&\quad= 1 + \left(\frac{1-\theta\cot\theta}{4\theta^2}\right)\tau + \order{\tau^2}.
\end{align}
In the expression \eqref{Eq MPSD correction}, this expansion was calculated out to $\order{\tau^4}$ and the coefficient functions of $\theta$ were expressed as truncated power series using the counting $\theta^2\sim \order{\tau}$.

\subsection{Boundary-layer expansion}\label{App BL}

When $\pi-\theta=O(\tau)$, the upper limit of the $v$ integral, $\theta(\pi-\theta)/(2\tau)\sim\pi(\pi-\theta)/(2\tau)$, remains finite as $\tau$ goes to zero and cannot be extended to infinity. Furthermore, in this same limit the $n=-1$ term in \eqref{Eq hk S2 exact} becomes comparable to the $n=0$ term.

Keeping both $n=0, -1$ terms and switching to variables suggested by the upper limit,
\begin{align}
	\xi \equiv \frac{\pi(\pi-\phi)}{2\tau}, \quad \eta \equiv \frac{\pi(\pi-\theta)}{2\tau},
\end{align}
we put the heat kernel in the form \eqref{Eq boundary layer},
\begin{gather}
K_{S^2}(\theta,\tau)\approx K_{BL}(\theta,\tau)= \frac{\sqrt{\pi}\,e^{\tau/4}}{4\tau^{3/2}}e^{-\frac{\pi^2}{4\tau}}G(\eta(\theta,\tau),\tau),\\
G(\eta,\tau)=\frac{2}{\pi^2}\int_0^\eta d\xi \frac{\pi\cosh\xi - \frac{2\tau}{\pi}\xi\sinh\xi}{\sqrt{2\left(\frac{\pi}{2\tau}\right)^2\left(\cos\left(\frac{2\tau\xi}{\pi}\right)-\cos\left(\frac{2\tau\eta}{\pi}\right)\right)}}e^{-\tau\frac{\xi^2}{\pi^2}}. \label{Def G}
\end{gather}
The integrand of $G$ is expanded in powers of $\tau$ keeping $\eta$ fixed. To $\order{\tau}$, the cosines in the denominator are expanded to second order,
\begin{align}
	G(\eta, \tau)&=\frac{2}{\pi^2}\int^\eta_0 \frac{d\xi}{\sqrt{\eta^2-\xi^2}}\left(\pi \cosh \xi - \frac{\tau}{\pi}\left(\xi^2 \cosh \xi + 2\xi \sinh \xi \right)+\order{\tau^2}\right)\non
	&= 	I_0(\eta)-\frac{\tau}{\pi^2}\left(\eta^2I_0(\eta)+\eta I_1(\eta)\right)+\order{\tau^2}.
\end{align}

To evaluate $G$ exactly at $\theta=\pi$, we change variables to $s=\xi/\eta$ in \eqref{Def G} before taking the limit $\eta\to 0$. This gives \begin{equation} G(0,\tau) = \frac{2}{\pi}\int_0^1\frac{ds}{\sqrt{1-s^2}} =1, \end{equation} in agreement with the known short-time asymptotics at the antipode \cite{hsu2002,barilariJendrej2014}.

\section{Tangent score representations: $S^2$ vs $S^3$ architectures}
\label{App score parametrization}

\subsection{Invariant vector fields and choice of architecture}

We consider the double covering $S^3\simeq SU(2)\to SO(3)$,
parametrized by unit quaternions $q=(q_0, q_1, q_2, q_3)$. Our convention for the corresponding
rotation matrix is
\begin{equation}
	R(q)=
	\begin{pmatrix}
		2(q_0^2+q_3^2)-1 & 2(q_0q_1-q_2q_3) & 2(q_0q_2+q_1q_3)\\
		-2(q_0q_1+q_2q_3) & 2(q_0^2+q_2^2)-1 & 2(q_0q_3-q_1q_2)\\
		2(q_1q_3-q_0q_2) & -2(q_0q_3+q_1q_2) & 1-2(q_2^2+q_3^2)
	\end{pmatrix}.\label{Eq R from q}
\end{equation}
This is antihomomorphic, $R(q q')=R(q')R(q)$.

A projection $\pi: SO(3) \rightarrow S^2$ is defined by $\pi(R) = R\hat{\mathbf r}_3 \equiv \mathbf{n}$, with $\hat{\mathbf r}_a$ the standard Cartesian basis. This is consistent with the definition of unit complex vector $z$ in the $CP^1$ model,
\begin{align}
	z = (q_0 + i q_1, q_2 + iq_3)^T,\quad \mathbf{n}= z^\dagger \boldsymbol{\sigma}z. \label{Def z from q}
\end{align}

Let $\Omega_a$ denote the three generators of $\mathfrak{so}(3)$,
defined by $\Omega_a\mathbf v=\hat{\mathbf{ r}}_a\times\mathbf v$. These are associated to right-invariant vector fields, i.e. the tangent vector to the path $e^{t\Omega_a}R$ at the point $R\in SO(3)$. The pushforward of the right-invariant vector field is
\begin{align}
	\mathbf{e}^R_a\equiv  \left.\frac{d}{dt}\pi(e^{t\Omega_a}R)\right|_{t=0}
	=\hat{\mathbf{ r}}_a\times\mathbf n.
\end{align}

In our implementation, the neural net generates the Lie algebra element $\omega^a_\theta \Omega_a$. The score $\mathbf{s}_\theta$ is constructed from the pushforward of the associated right-invariant
vector field,
\begin{align}
	\mathbf{s}_\theta
=\sum_a\omega_\theta^a\mathbf{e}^R_a
=\boldsymbol{\omega}_\theta\times\mathbf{n}.\label{Score S2}
\end{align}

Alternatively, one might generate the score from
a \emph{left}-invariant vector field on $SO(3)$,
\begin{equation}
 \mathbf{e}^L_a \equiv	\left.\frac{d}{dt}\pi(Re^{t\Omega_a})\right|_{t=0}
=R(\hat{\mathbf r}_a\times\hat{\mathbf r}_3).
\end{equation}
Explicitly, this basis is given by the first and second columns of $R$ in \eqref{Eq R from q},
\begin{align}
\mathbf e_1^L=-R\hat{\mathbf r}_2, \quad 
\mathbf e_2^L= R\hat{\mathbf r}_1, \quad \mathbf e_3^L = 0. \label{Def eL}
\end{align}

A neural net outputs coefficients of the left-invariant fields, giving score
\begin{align}
	\mathbf{s}_\theta
	= \omega_\theta^1\mathbf{e}^L_1 + \omega_\theta^2\mathbf{e}^L_2. \label{Score S3}
\end{align}
The component $\omega^3_\theta$ would be redundant, so the neural net need only generate two components. But the basis $\mathbf{e}_a^L$ depends on all columns of $R$, not just $\mathbf{n}$, so the model must receive extra gauge input which is not fixed by $\mathbf{n}$ alone. To implement this parameterization, we use random representatives of the $CP^1$ field $z$, or equivalently random quaternions $q\in S^3$, as inputs to the neural network.

\begin{figure}
	\centering
	\includegraphics[width=0.6\textwidth]{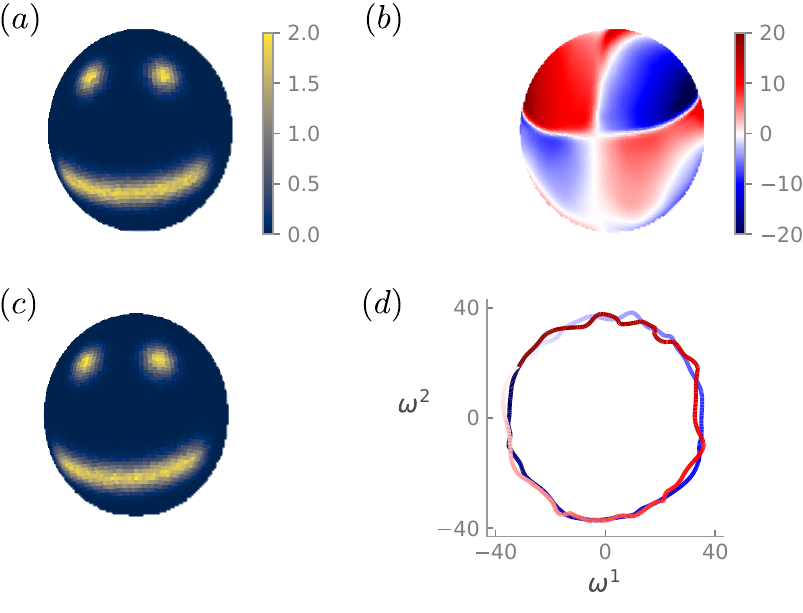}
	\caption{A comparison of two architectures for learning a smiley face-shaped trimodal distribution on $S^2$. We test both a model taking direct $S^2$ inputs (a,b) and a model taking inputs randomly lifted to $S^3$ (c,d). Histograms of reverse diffusion samples from the respective models are shown in (a) and (c). The component $\boldsymbol{\omega}_\theta\cdot\mathbf{n}$, which is unconstrained by the loss, is shown in (b). For a fixed point on $S^2$, the dependence of $\omega_\theta^{1,2}$ on the gauge phase (indicated by color) is shown in (d). The nearly doubly wound circle demonstrates approximate gauge equivariance.}\label{Fig smiley}
\end{figure}

\subsection{Numerical comparison of the $S^2$ and $S^3$ networks}

We refer to the architectures defined by \eqref{Score S2} and \eqref{Score S3} as the $S^2$ and $S^3$ networks, respectively. We test these choices of architecture by training them on a target distribution defined on a single copy of $S^2$, see Fig.~\ref{Fig smiley}. Both models use multilayer perceptrons (MLPs) with four hidden layers of width 256. The $S^2$ model takes $\mathbf{n}, u$ as input and outputs the three-component vector $\boldsymbol{\omega}_\theta$. The $S^3$ model takes $q, u$ as input and outputs the two components $\omega^1_\theta, \omega^2_\theta$.

To construct a random quaternion representative above a given spin $\mathbf n=(n_1, n_2, n_3)$, we first choose a reference representative. For the northern hemisphere, setting $q_1=0$ and $q_0\geq0$ in \eqref{Eq R from q} gives
\begin{align}
	q_0=\sqrt{\frac{1+n_3}{2}},\qquad
	q_2=\frac{n_1}{2q_0},\qquad
	q_3=\frac{n_2}{2q_0}.
\end{align}
We then apply a random phase $\alpha$ to the corresponding complex unit vector, $z\to e^{i\alpha}z$, obtaining a uniform representative of the $U(1)$ fiber. The southern hemisphere is treated analogously using a reference representative regular at the south pole.

The smiley face-shaped target distribution is constructed as a sum of spherical heat kernels with diffusion time $\tau_0=0.005$, the same parameter used in our definition of $u(\tau)$. Under subsequent forward diffusion, the distribution remains a sum of heat kernels with diffusion time $\tau_0+\tau$, so its score is analytically known for all $u$. For this test, we train the models directly on the analytic score rather than using DSM on either $S^2$ or $S^3$.

To quantify the error in diffusion model sampling, we construct a histogram involving the generated samples and compute its total variation (TV) distance from the exact distribution, integrated over the same bins. Dividing the sphere into 8,192 equal area bins, a histogram of $10^5$ samples of the exact distribution has TV distance of $0.0387(6)$, whereas the corresponding TV distances of samples from the $S^2$ and $S^3$ models, using 1600 steps of Euler-Maruyama, are $0.0381$ and $0.0403$ respectively (Fig.~\ref{Fig smiley}(a,c)).

For the $S^2$ model, the redundant component $\boldsymbol{\omega}_\theta\cdot\mathbf{n}$ at $u=0$ is shown in Fig.~\ref{Fig smiley}(b). This redundancy does not appear to cause problems in practice, and we use the $S^2$ architecture for the $O(3)$ model experiments in the main text. In contrast, the $S^3$ model has no redundant output component, but introduces gauge redundancy in the input. As the gauge phase $\alpha$ in $e^{i\alpha}z$ varies from $0$ to $2\pi$, the output vector $(\omega_\theta^1,\omega_\theta^2)$ should undergo a rotation of $4\pi$. Figure~\ref{Fig smiley}(d) shows that this gauge equivariance is approximately, but not exactly, learned by the network.

\subsection{Gauge-equivariant architectures}

We say that a quantity has charge $n$ if it transforms by a factor $e^{in\alpha}$ under a gauge rotation by $\alpha$. The input $z$ has charge $1$ whereas the output $\omega^1_\theta+i\omega^2_\theta$ has charge $2$. We briefly consider some approaches to enforcing this gauge equivariance exactly.

A simple but computationally expensive way to improve the gauge equivariance of the $S^3$ architecture is to perform multiple full score evaluations along the gauge fiber, which are used in a discrete Fourier integral to project on the charge $2$ component. A more sophisticated implementation of the same idea is developed in \cite{deHaanEtAl2021,franzenWand2021}, where internal features are represented by components of definite charge, and Fourier projection is used after nonlinear operations.

An alternative approach uses only gauge-invariant internal features. Note that we can form charge $2$ combinations from products of the $z_0=q_0+iq_1$ and $z_1 = q_2+iq_3$ components of the input. These may be expressed as linear combinations of the components of the $\mathbf{e}^L_+ \equiv  \mathbf{e}^L_1 + i\mathbf{e}^L_2$ vector. So a possible gauge-equivariant architecture outputs a complex three-vector $\mathbf{f}+i \mathbf{g}$ as a function of gauge-invariant $\mathbf{n}$,
\begin{align}
\omega^1_\theta+i\omega^2_\theta = \left(\mathbf{f}_\theta+i\mathbf{g}_\theta \right)\cdot \mathbf{e}^L_+.
\end{align}
Given the relation for the score \eqref{Score S3}, this is equivalent to
\begin{align}
	\mathbf{s}_\theta = \left(\mathbf{n}\times\mathbf{f}_\theta -\mathbf{g}_\theta  \right)\times \mathbf{n}.
\end{align}
This is just a slight modification of the $S^2$ architecture, with $\boldsymbol{\omega}_\theta =  \mathbf{n}\times\mathbf{f}_\theta -\mathbf{g}_\theta $, and as discussed in connection with \eqref{Eq score overcomplete}, we may set one of $\mathbf{f}$ or $\mathbf{g}$ equal to zero with no loss of generality.

One possible benefit of this lift of the $S^2$ architecture to a gauge-equivariant $S^3$ architecture is that it allows the use of heat kernels on $S^3$ or $SO(3)$ in the DSM loss, which admit simpler short-time approximations, as in \eqref{Eq hk SO3}. The network output $\omega^1_\theta,\omega^2_\theta$ is interpreted as components of a tangent vector on $SO(3)$ in the left-invariant basis, with the component $\omega^3$ set to zero a priori. This is matched to the conditional score of the $SO(3)$ heat kernel. The DSM procedure on $SO(3)$ involves randomly lifting each initial spin $\mathbf{n}_0$ to $SO(3)$ by choosing a uniform gauge angle $\alpha$. Using the fiber-integral identity \eqref{Eq hk S2 fiber} and the invariance of the $SO(3)$ heat kernel under simultaneous rotations of its arguments, one can show that the expected DSM loss is equivalent to that obtained using the $S^2$ heat kernel, up to a term independent of the network parameters.

\subsection{Generalization to $CP^{N-1}$}

We briefly discuss the natural extension of the $S^2$ and $S^3$ architectures to the $CP^{N-1}$ family of manifolds. Here $z$ is a charge-$1$ complex unit vector with $N$ components. Generalizing the $S^3$ architecture, a model may take $z$ as input and output a charge-$1$ horizontal tangent vector $v_\theta$ satisfying $z^\dagger v_\theta=0$. For $N=2$, this vector can be written as $v_\theta=w_\theta\tilde z$, where $w_\theta\equiv\omega_\theta^1+i\omega_\theta^2$ and $\tilde z\equiv(-z_1^*,z_0^*)^T$. For $N>2$, there is no unique extension of $\tilde z$, and the charge-$2$ quantity $w$ generalizes naturally to the antisymmetric tensor $w_{ij}=z_i v_j-z_j v_i$. 
%

The $S^2$ architecture generalizes to $CP^{N-1}$ through the adjoint representation of $SU(N)$. The gauge-invariant coordinates $n^a=2z^\dagger T^a z$, where $T^a$ are the Hermitian generators of $SU(N)$, generalize the relation $n^a=z^\dagger\sigma^a z$ for $N=2$. These $N^2-1$ coordinates lie on a submanifold defined by constraints inherited from the projector condition $P^2=P$, where $P=zz^\dagger$. A network taking $n^a$ as input may output $N^2-1$ components $\omega_\theta^a$, from which a tangent score is constructed as
\begin{align}
	s_\theta^a=f^{abc}\omega_\theta^b n^c,
\end{align}
where $f^{abc}$ are the structure constants of $SU(N)$. This is the natural generalization of $\mathbf{s}_\theta=\boldsymbol{\omega}_\theta\times\mathbf{n}$, obtained by pushing forward the invariant vector fields of $SU(N)$ through its adjoint action. 

\section{Gaussian interpolation}\label{App Interp}

\begin{figure}
	\centering
	\includegraphics[width=0.8\textwidth]{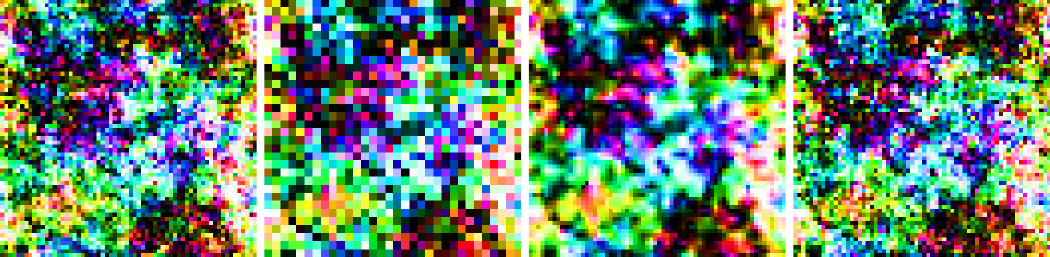}
	\caption{An illustration of various interpolation schemes. The original configuration is shown at left. This is coarse-grained and then interpolated back to the original lattice size using block replication (center-left), linear interpolation from neighboring coarse spins (center-right), and Gaussian interpolation (right).}\label{Fig interp}
\end{figure}

The prolongation steps in the superresolution methods discussed here use an interpolation scheme based on that of Hoffman and Ribak \cite{hoffmanRibak1991}. The spins on a coarse sublattice are held fixed, while the remaining spins are sampled from a Gaussian distribution designed to reproduce the measured short-distance correlations.

Consider a Gaussian distribution for schematic variables $x$, which in our context represent the spin components $n^A_i$. The quadratic form in the exponent of the Gaussian distribution is $x^T K^{-1} x$, where $K$ is the covariance matrix of the variables $x$, and is concretely given by the two-point correlation function $\langle n^A_i n^B_j\rangle$ calculated in the training set. Distinguishing coarse and fine variables $x_C,x_F$, we decompose $K$ into blocks as
\begin{gather}
	K = \left(\begin{array}{cc}
		K_{CC} & K_{CF}\\
		K_{FC} & K_{FF}
	\end{array}\right)
	= \left(\begin{array}{cc}
		I & 0\\
		W & I
	\end{array}\right)
	\left(\begin{array}{cc}
		K_{CC} & 0\\
		0 & \Sigma
	\end{array}\right)
	\left(\begin{array}{cc}
		I & W^T\\
		0 & I
	\end{array}\right),\\
	W \equiv K_{FC}K_{CC}^{-1},\qquad
	\Sigma \equiv K_{FF}-K_{FC}K_{CC}^{-1}K_{CF}.
\end{gather}
The quadratic form then becomes
\begin{align}
	x^T K^{-1} x =
	x_C^T K_{CC}^{-1}x_C
	+ (x_F-Wx_C)^T\Sigma^{-1}(x_F-Wx_C).
\end{align}
Thus, holding $x_C$ fixed, the conditional distribution of $x_F$ is Gaussian with mean $Wx_C$ and covariance $\Sigma$.

This allows us to sample the fine spins $x_F$ directly from a multivariate Gaussian with these parameters. However, a more efficient approach is to first sample variables $z_C,z_F$ from a zero-mean Gaussian with unconditional covariance $K$, and construct
\begin{align}
	x_F = z_F + W(x_C-z_C).
\end{align}
By direct calculation, this has the required conditional mean and covariance.

The vectors $\mathbf{n}_i$ constructed at fine lattice sites by this method do not generally satisfy the normalization condition $|\mathbf{n}_i|^2=1$. When refinement is performed using heat-bath sweeps, this is not problematic, since the unnormalized spins are used only in the nearest-neighbor field $\mathbf{m}$ in the initial sweep. However, in the method discussed in Sec.~\ref{Sec superres}, where refinement uses Langevin corrections based on the learned score, the spins must be normalized after interpolation. This spoils the exact reproduction of the prescribed Gaussian covariance.

Nevertheless, Gaussian interpolation followed by normalization produces better results than several simpler alternatives, including block replication and local linear interpolation followed by normalization. This is illustrated visually in Fig.~\ref{Fig interp}.

\section{Equilibration with heat-bath Monte Carlo}\label{App equilibration}

\begin{figure}
	\centering
	\includegraphics[width=\textwidth]{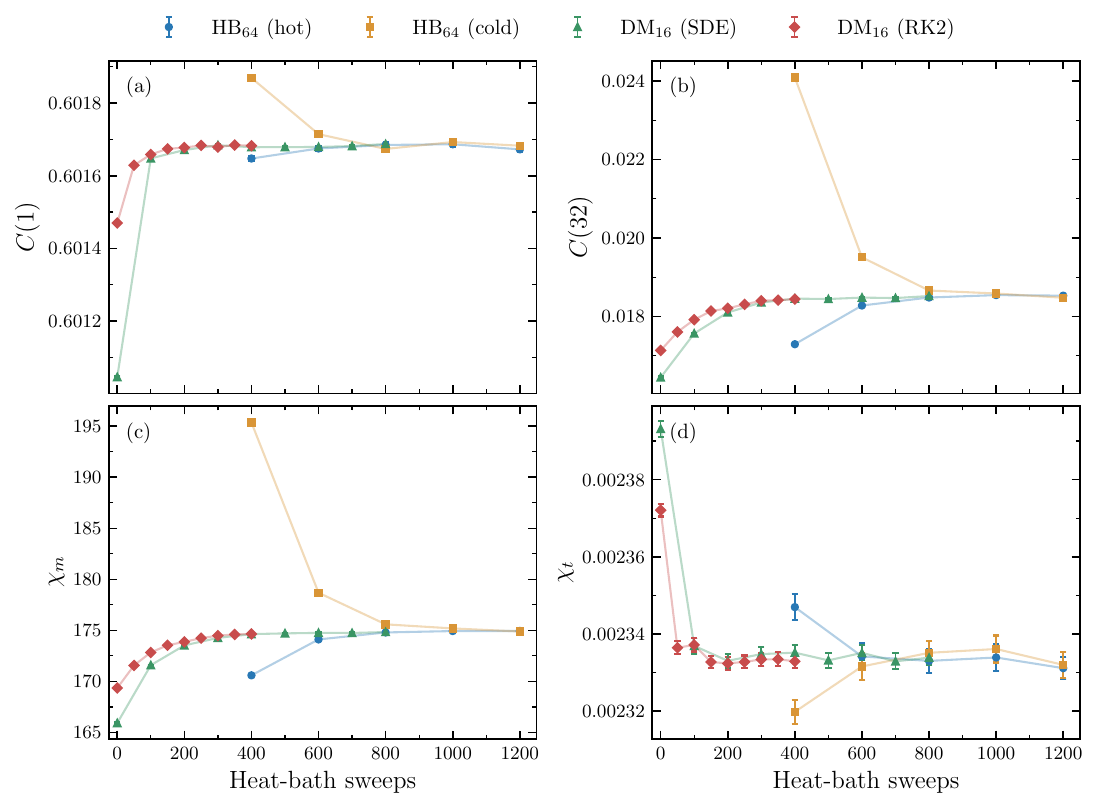}
	\caption{Equilibration under heat-bath Monte Carlo on the $64^2$ lattice, starting from random (hot), uniform (cold), and diffusion-generated configurations. The four panels show the nearest-neighbor correlation $C(1)$, long-distance correlation $C(32)$, magnetic susceptibility $\chi_m$, and topological susceptibility $\chi_t$. Diffusion-generated configurations are obtained using $\mathrm{DM}_{16}$ with RK2 or mixed SDE sampling, followed by prolongation and restricted heat-bath refinement. For these curves, zero sweeps corresponds to the beginning of unrestricted heat-bath evolution, after prolongation and restricted refinement have been completed.}\label{Fig equilibration}
\end{figure}

We investigate whether configurations generated using $\mathrm{DM}_{16}$ can accelerate equilibration under heat-bath Monte Carlo. We generated $4\times10^6$ configurations using RK2 integration \eqref{Eq ODE} with 100 reverse diffusion steps (200 score evaluations), and $3\times10^6$ configurations using a mixed SDE scheme with the same evaluation budget, including 100 Langevin correction steps \eqref{Eq langevin correction} with $dv=0.01$ at $u=0.5$.

The configurations are prolonged directly from $16^2$ to $64^2$ using conditional Gaussian interpolation \cite{hoffmanRibak1991}, followed by heat-bath refinement with the original coarse spins held fixed. Tests using decimated equilibrium configurations showed that approximately 40 restricted sweeps reproduced the original correlations within statistical uncertainty. For diffusion-generated configurations, we obtained better results using 20 restricted sweeps before releasing the constraints and continuing with ordinary heat-bath updates. Using the A100 timings in Table~\ref{Table computational cost}, the 200 $\mathrm{DM}_{16}$ score evaluations and 20 restricted heat-bath sweeps cost approximately 440 equivalent $\mathrm{HB}_{64}$ sweeps, excluding the comparatively inexpensive interpolation.

Figure~\ref{Fig equilibration} compares the subsequent equilibration with $10^6$ independent configurations initialized randomly (hot) or uniformly (cold). The nearest-neighbor correlation $C(1)$ and topological susceptibility $\chi_t$ approach equilibrium relatively quickly for all initial conditions, while the long-distance correlation $C(32)$ and magnetic susceptibility $\chi_m$ require substantially more heat-bath sweeps. Although the diffusion-initialized ensembles begin closer to equilibrium, accounting for the initialization cost makes their overall computational performance roughly comparable to conventional heat-bath sampling.

\providecommand{\href}[2]{#2}\begingroup\raggedright\endgroup

\end{document}